\documentclass[11pt,twocolumn]{article}

\usepackage[letterpaper,top=2cm,bottom=2.5cm,left=1.5cm,right=1.5cm,marginparwidth=1.75cm]{geometry}

\usepackage{comment,caption,amsbsy,subfig}
\usepackage{pdflscape,bigints,changepage}
\usepackage{float}
\usepackage{natbib}
\usepackage{amsmath, amsthm, amssymb,lipsum,xfrac}
\usepackage{graphicx, color, xcolor}
\usepackage{mathrsfs,enumitem}
\usepackage{authblk}
\newtheorem{theorem}{Theorem}%
\newtheorem{remark}{Remark}%

\newcommand\blfootnote[1]{%
  \begingroup
  \renewcommand\thefootnote{}\footnote{#1}%
  \addtocounter{footnote}{-1}%
  \endgroup
}

\usepackage{hyperref}
\hypersetup{
    colorlinks = true,
    linkcolor = blue,
    anchorcolor = blue,
    citecolor = blue,
    filecolor = blue,
    urlcolor = blue
    }
\title{\huge\sffamily Improving Model Choice in Classification: An Approach
Based on Clustering of Covariance Matrices.}
\author{}
\date{}

\begin{document}

\twocolumn[%
   \begin{center}
   \maketitle
   \vspace{-1.5cm}
      \large{
      \textsc{David Rodríguez-Vítores\footnotemark[1] \footnotemark[2] and Carlos Matrán \footnotemark[1] }
      \vspace{0.5cm}
      
       \textsc{\footnotemark[1] Department of Statistics and Operational Research and IMUVA,\\ University of Valladolid, Spain.}}
   \end{center}

   \begin{adjustwidth}{1cm}{1cm}
      \begin{abstract}
 This work introduces a refinement of the Parsimonious Model for fitting a Gaussian Mixture. The improvement is based on the consideration of {clusters of the involved} covariance matrices according to a criterion, such as sharing Principal Directions. This and other similarity criteria that arise from the spectral decomposition of a matrix are the bases of the Parsimonious Model. {We show that such groupings of covariance matrices can be achieved through simple modifications of the CEM (Classification Expectation Maximization) algorithm. Our} approach leads to propose Gaussian Mixture Models for model-based clustering and discriminant analysis, in which covariance matrices are clustered according to a parsimonious criterion, creating intermediate steps between the fourteen widely known parsimonious models. The added versatility not only allows us to obtain models with fewer parameters for fitting the data, but also provides greater interpretability. We show its usefulness for model-based clustering and discriminant analysis, providing algorithms to find approximate solutions verifying suitable size, shape and orientation constraints, and applying them to both simulation and real data examples.
       \end{abstract}
  \end{adjustwidth}

   \vspace{0.05cm}%
]
\footnotetext[2]{\textit{email:} david.rodriguez.vitores@uva.es}

\blfootnote{\noindent \vspace{0.01cm}

The research leading to these results received funding from MCIN/AEI/10.13039/501100011033/FEDER under Grant Agreement No PID2021-128314NB-I00}

\section{Introduction}\label{sec1}
\footnote{The research leading to these results received funding from MCIN/AEI/10.13039/501100011033/FEDER under Grant Agreement No PID2021-128314NB-I00}
In this paper we introduce methodological applications arising of cluster analysis of covariance matrices. Throughout, we will show that appropriate clustering criteria on these objects provide useful tools in the analysis of classic problems in Multivariate Analysis. The chosen framework is that of multivariate classification under a Gaussian \\

\vspace{0.8cm} 
Mixture Model, a setting where a suitable reduction of the involved parameters is a fundamental goal leading to the Parsimonious Model. We focus on this hierarchized model, designed to explain data with a minimum number of parameters, by introducing intermediate categories associated with clusters of covariance matrices. \\

Gaussian Mixture Models approaches to discriminant and cluster analysis are well-established and powerful tools in multivariate statistics. For a fixed number ${{K}}$, both methods aim to fit $ {{K}}$ multivariate Gaussian distributed components to a data set in $\mathbb R^d$,  with the key difference that labels providing the source group of the data are known (supervised classification) or unknown (unsupervised classification).
In the supervised problem, we handle a data set with $N$ observations $y_1,\ldots,y_N$ on $\mathbb R^d$ and associated labels $z_{{i},{{k}}},{i}=1,\ldots,N$, ${{k}}=1,\ldots,{K}$, where $z_{{i},{{k}}}=1$ if the observation $y_{{i}}$ belongs to the group ${k}$ and 0 otherwise. Denoting by $\phi(\cdot \vert \mu,\Sigma)$ the density of a multivariate Gaussian distribution on $\mathbb R^d$ with mean $\mu$ and covariance matrix $\Sigma$, we seek to maximize the complete log-likelihood function
\begin{equation} \label{eq:cloglik}
CL\Bigl(\pmb\pi,\pmb\mu,\pmb\Sigma\Bigr) = \sum_{{{i}}=1}^N\sum_{{k}=1}^{{K}} z_{{i},{{k}}} \log\Biggl( \pi_{{k}} \phi(y_{{i}}\vert\mu_{{k}},\Sigma_{{k}})\Biggr) ,
\end{equation}
with respect to the weights $\pmb{\pi}=(\pi_1,\ldots,\pi_{{K}})$ with $0 \leq\pi_{{k}}\leq 1, \  \sum_{{k}=1}^{{K}} \pi_{{k}}=1$, the means $\pmb{\mu}=(\mu_1,\ldots,\mu_{{K}})$ and the covariance matrices $\pmb{\Sigma}=(\Sigma_1,\ldots,\Sigma_{{K}})$. In the unsupervised problem the labels $z_{{i},{{k}}}$ are unknown, and fitting the model involves the maximization of the log-likelihood function
\begin{equation} \label{eq:loglik}
  L\Bigl( \pmb\pi,\pmb\mu,\pmb\Sigma  \Bigr) = \sum_{{{i}}=1}^N\log\Biggl(\sum_{{k}=1}^{{K}} \pi_{{k}} \phi(y_{{i}}\vert \mu_{{k}},\Sigma_{{k}})\Biggr) \ ,
\end{equation}
with respect to the same parameters. This maximization is more complex, and it is usually performed via the EM algorithm \citep{EM}, where we repeat iteratively the following two steps. The E step, which consists in computing the expected values of the unobserved variables $z_{{i},{{k}}}$ given the current parameters, and the M step, in which we are looking for the parameters maximizing the complete log-likelihood (\ref{eq:cloglik}) for the values $z_{{i},{{k}}}$ computed in the E step. Therefore, both model-based techniques require the maximization of (\ref{eq:cloglik}), for which optimal values of the weights and the mean are easily computed:
\begin{equation}\label{eq:optValues}
        n_{{k}}= \sum_{{{i}}=1}^N z_{{i},{{k}}}\ , \ \pi_{{k}}=\frac{n_{{k}}}{N}\ , \  \mu_{{k}}
        =\frac{\sum_{{{i}}=1}^N z_{{i},{{k}}}\thinspace y_{{i}}}{n_{{k}}} \ .
    \end{equation}
With these optimal values, if we denote $S_{{k}}=(1/n_{{k}})\sum_{{{i}}=1}^N z_{{i},{{k}}} (y_{{i}}-\mu_{{k}})(y_{{i}}-\mu_{{k}})^T$, the problem of maximizing (\ref{eq:cloglik}) with respect to $\Sigma_1,\ldots,\Sigma_{{K}}$ is equivalent to the problem of maximizing
\begin{equation} \label{eq:intr0}
   (\Sigma_1,\ldots,\Sigma_{{K}}) \mapsto \sum_{{k}=1}^{{K}} \hspace{0.2cm} \log\Bigl(W_d\bigl(n_{{k}}S_{{k}}\vert n_{{k}},\Sigma_{{k}}\bigr)\Bigr) \ ,
\end{equation}
where $ W_d(\thinspace \cdot \thinspace \vert n_{{k}},\Sigma_{{k}})$ is the ${d}$-dimensional Wishart distribution with parameters $n_{{k}},\Sigma_{{k}}$. For even moderate dimension $d$, the large number of involved parameters in relation with the size of the data set could result in a poor behavior of standard unrestricted methods. In order to improve the solutions, regularization techniques are often invoked. In particular, many authors have proposed estimating the maximum likelihood parameters under some additional constraints on the covariance matrices $\Sigma_1,\ldots,\Sigma_{{K}}$, which lead us to solve the maximization of (\ref{eq:intr0}) under these constraints. Between these proposals, a  prominent place is occupied by the so called \textbf{Parsimonious Model}, a broad set of hierarchized constraints capable of adapting to conceptual situations that may occur in practice. \\

 A common practice in multivariate statistics consists in assuming that covariance matrices share a common part of their structure. For example, if $\Sigma_1=\ldots=\Sigma_{{K}}=I_d$, the clustering method described in {(\ref{eq:loglik})} gives just the k-means. If we assume common covariance matrices $\Sigma_1=\ldots=\Sigma_{{K}}=\Sigma$, the procedure coincides with linear discriminant analysis (LDA) in the supervised case {(\ref{eq:cloglik})}, and with the method proposed in \cite{friedman} in the unsupervised case {(\ref{eq:loglik})}. General theory to organize these relationships between covariance matrices is based on the spectral decomposition, beginning with the analysis of Common Principal Components \citep{Flury,Flury2}. In the discriminant analysis setting, the use of the spectral decomposition was first proposed in \cite{FluryDisc}, and in the clustering setting in \cite{Raftery}. The term ``Parsimonious model" and the fourteen levels given in Table \ref{tabla:parsimoniosos} were introduced in \cite{Parsimonious} for the clustering setting and later, in \cite{discriminante2}, for the discriminant setup.
 \begin{table*}[ht]
     \caption{Parsimonious levels based on the spectral decomposition of $\Sigma_1,\ldots,\Sigma_{{K}}.$}
     \medskip
    \centering
    \begin{tabular}{cccccc}
    \hline
    \hspace{0.3cm}Name \hspace{0.3cm}  & 
      \hspace{0.3cm}$\Sigma_{{k}}$\hspace{0.3cm}  &  
      \hspace{0.3cm}Size\hspace{0.3cm} & \hspace{0.3cm}Shape\hspace{0.3cm} & \hspace{0.3cm}Orientation\hspace{0.3cm} & \hspace{0.3cm}Parameters\hspace{0.6cm} \\
      \hline
      EII &${\gamma}I$ & Equal & Spherical & - & ${1}$ \\
      VII & ${\gamma}_{{k}} I$ & Variable & Spherical & - & ${K}$ \\
      EEI& ${\gamma}\Lambda$ & Equal & Equal & Canonical & ${1 + (d-1)}$ \\
       EVI & ${\gamma}\Lambda_{{k}}$ & Equal & Variable & Canonical & ${1+K(d-1)}$\\
      VEI& ${\gamma}_{{k}} \Lambda$ & Variable & Equal & Canonical & ${ K + (d-1)}$ \\
       VVI& ${\gamma}_{{k}} \Lambda_{{k}}$ & Variable & Variable & Canonical &  ${K + K(d-1)} $\\
       EEE & ${\gamma}\beta \Lambda \beta^T$ & Equal & Equal & Equal & ${1+(d-1)+d(d-1)/2} $\\
        EEV & ${\gamma}\beta_{{k}} \Lambda \beta_{{k}}^T$ & Equal & Equal & Variable & $ {1+(d-1)+Kd(d-1)/2} $ \\
         EVE& ${\gamma}\beta \Lambda_{{k}} \beta^T$ & Equal & Variable & Equal &  ${1+K(d-1)+d(d-1)/2}$  \\
        VEE & ${\gamma}_{{k}} \beta \Lambda \beta^T$ & Variable & Equal & Equal &  ${K+(d-1)+d(d-1)/2}$ \\
        VVE &  ${\gamma}_{{k}} \beta \Lambda_{{k}} \beta^T$ & Variable & Variable & Equal & $ {K+K(d-1)+d(d-1)/2}$  \\
        EVV &  ${\gamma}\beta_{{k}} \Lambda_{{k}} \beta_{{k}}^T$ & Equal & Variable & Variable &  ${1+K(d-1)+Kd(d-1)/2 }$\\
          VEV&  ${\gamma}_{{k}} \beta_{{k}} \Lambda \beta_{{k}}^T$ & Variable & Equal & Variable &  ${K+(d-1)+Kd(d-1)/2}$  \\
         VVV&  ${\gamma}_{{k}} \beta_{{k}} \Lambda_{{k}} \beta_{{k}}^T$ & Variable & Variable & Variable & $ {K(1+(d-1)+d(d-1)/2)}$ \\
          \hline   
    \end{tabular}
    \label{tabla:parsimoniosos}
\end{table*}
Given a positive definite covariance matrix $\Sigma_{{k}}$, the spectral decomposition of reference is
\begin{equation*}\Sigma_{{k}} = \gamma_{{k}} \beta_{{k}} \Lambda_{{k}} \beta_{{k}}^T \ ,\end{equation*}
where $\gamma_{{k}}=\operatorname{det}(\Sigma_{{k}})^{1/d}> 0$ governs the size of the groups, $\Lambda_{{k}}$ is a diagonal matrix with positive entries and determinant equal to 1 that controls the shape, and $\beta_{{k}}$ is an orthogonal matrix that controls the orientation. Given ${{K}}$ covariance matrices $\Sigma_1,\ldots,\Sigma_{{K}}$, the spectral decomposition enables to establish the fourteen different parsimonious levels in Table \ref{tabla:parsimoniosos}, allowing differences or not in the parameters associated to size, shape and orientation.
  To fit a Gaussian Mixture Model under a parsimonious level $\mathscr M$ in the Table \ref{tabla:parsimoniosos}, we must face the maximization of (\ref{eq:intr0}) under the parsimonious restriction. That is, we should find 
    \begin{align} \label{eq:intr1}
    \hat{\pmb \Sigma}=
\underset{{\pmb \Sigma }\in \mathscr M}{\operatorname{argmax}}\  \sum_{{k}=1}^{{K}} \hspace{0.1cm} \log\Bigl(W_d\bigl(n_{{k}}S_{{k}}\vert n_{{k}},\Sigma_{{k}}\bigr)\Bigr) ,
\end{align}
where we say that ${\pmb \Sigma }=(\Sigma_1,\ldots,\Sigma_{{K}})\in\mathscr M$ if the ${K}$ covariance matrices  verify the level. We should remark that  the Common Principal Components model \citep{Flury,Flury2} plays a key role in this   hierarchy, which in any case is based on simple geometric interpretations.\\
    
Restrictions are also often used to solve a well-known problem that appears in model-based clustering, the unboundedness of the log-likelihood function (\ref{eq:loglik}). With no additional constraints, the problem of maximizing (\ref{eq:loglik}) is not even well defined, a fact that could lead to uninteresting spurious
solutions, where some groups would be associated to a few, almost collinear, observations. Although we will also use these restrictions, we will not discuss on this line in this work. A review of approaches for dealing with this problem can be found in \cite{eigConstr}. \\

The aim of this paper is to introduce a generalization of equation (\ref{eq:intr1}), that allows us to give a likelihood-based classification {  associated to intermediate parsimonious levels. Let $G \in \{1,\ldots,K\}$ and $\pmb u=(u_1,\ldots,u_K)$ be any vector in $\lbrace 1,\ldots, G \rbrace^{K}$. Given a parsimonious level $\mathscr M$, we can formulate a model in which we assume that the theoretical covariance matrices $\Sigma_1,\ldots,\Sigma_K$ verify a parsimonious level $\mathscr M$ within each of the $G$ classes defined by $\pmb u$. For instance, let $K=7$, $G=3$,  $\mathscr M$ = VVE and take $\pmb u=(1,1,2,3,1,2,1)$.  This implies
\begin{align*}
       \Sigma_{{k}} &= \gamma_{{k}}\beta_1\Lambda_{{k}} \beta_1^T, \quad {k}=1,2,5,7, \\
        \Sigma_{{k}} &= \gamma_{{k}}\beta_2\Lambda_{{k}} \beta_2^T, \quad {k}=3,6, \\
       \Sigma_{{k}} &= \gamma_{{k}}\beta_3\Lambda_{{k}} \beta_3^T ,\quad {k}=4 \ . 
\end{align*}
Following (\ref{eq:intr1}), the estimation of the original covariance matrices involves maximizing (\ref{eq:intr0}) within $\mathscr M_{\pmb u}$, the set of covariance matrices satisfying $\{\Sigma_k: u_k=g\} \in \mathscr M$ for all $g=1,\ldots,G$. Using the maximized log-likelihood as a measure for the appropriateness of $\pmb u$, the optimal $\hat{\pmb u}$ would provide a classification for $S_1,\ldots,S_K$ according to the level $\mathscr M$. Precise definitions will be provided in Section \ref{section:clas}. We will present an iterative procedure to simultaneously compute the optimal classification and covariance matrix estimators through the  modification of equation (\ref{eq:intr1}) given by
\begin{align} \label{eq:intr2}
 \bigl(&\mathbf{\hat{u}}, \mathbf{\hat{\Sigma}} \bigr)= \\ &\underset{\pmb u , \pmb \Sigma \in \mathscr M_{\pmb u}}{\operatorname{argmax}} \Biggl(\sum_{g=1}^G    \sum_{k:u_k=g}\log\Bigl(W_d(n_k S_k\vert n_k,\Sigma_k)\Bigr)\Biggr) .\nonumber
\end{align}
}

Solving this equation will allow us to fit Gaussian Mixture Models with intermediate parsimonious levels, in which the common parameters of a parsimonious level will be shared within each of the $G$ classes given by the vector of indexes ${\pmb{\hat{u}}}$, but varying between the different classes.  {In the previous example, we obtain three} classes of covariance matrices that share their principal directions within each class, resulting in a better interpretation of the final classification and allowing a considerable reduction of the number of parameters to be estimated. We will use these ideas for fitting Gaussian Mixture Models in discriminant analysis and cluster analysis. {To avoid unboundedness of the objective function in the clustering framework, we will impose the determinant and shape constraints of \cite{detShape}, which are fully implemented in the MATLAB toolbox FSDA \citep{FSDA}}. We will analyze some examples where the proposed models result in less parameters and more interpretability fitting the data, being better suited when compared with the 14 parsimonious models.  {We point out that, as it is becoming usual in the literature, to carry out the comparisons between different models,  we will use the Bayesian Information Criterion (BIC). This applies to all examples considered in the text.} It has been noticed by many authors that BIC selection works properly in model based clustering, as well as in discriminant analysis. \cite{mclust1} includes a detailed justification for the use of BIC, based on previous references. A summary of the comparison of BIC with other techniques for model selection can also be found in \cite{BIC}.  \\

The paper is organized as follows. Section \ref{section:clas} approaches the problem of the parsimonious classification of covariance matrices given by equation (\ref{eq:intr2}), focusing on its computation for the most interesting restrictions in terms of dimensionality reduction and interpretability. Throughout, we will only work  with models based on the parsimonious levels of proportionality (VEE) and common principal components (VVE), although the extension to other levels is straightforward. Section \ref{section:GMM} applies the previous theory for the estimation of Gaussian Mixture Models in cluster analysis and discriminant analysis, including some simulation examples for their illustration. Section \ref{section:examples} includes real data examples, where we will see the gain in interpretability that can arise from these solutions.  Some conclusions are outlined in Section \ref{sec:conclusion}. { Finally, Appendix \ref{app:teor} includes theoretical results, Appendix \ref{app:additionalSim} provides some additional simulation examples and Appendix \ref{appendixB} explains technical details about the algorithms.} Additional graphical material  is provided in the Online Supplementary Figures document.

\section{Parsimonious Classification of Covariance Matrices} \label{section:clas}

Given $n_1,\ldots,n_{{K}}$ independent observations from ${{K}}$ groups with different distributions, and $S_1,\ldots,S_{{K}}$ the sample covariance matrices, a group classification may be provided according to different similarity criteria. In the general case, given a similarity criterion $f$ depending on  the sample covariance matrices and the sample lengths, the problem of classifying ${{K}}$ covariance matrices in $G$ classes, $1\leq G \leq {{K}}$, typically would consist in solving the equation
\begin{equation*}
    {\hat{{\pmb u}}}= \underset{{\pmb u} \in \mathscr H}{\textnormal{argmax}} \quad \sum_{g=1}^G f\Bigl( \hspace{0.05cm} \bigl\lbrace (S_{{k}},n_{{k}}) : {u}_{{k}}= g \bigr\rbrace \Bigr) ,
\end{equation*}
where $\mathscr H = \bigl \lbrace{\pmb u}=(u_1,\ldots,u_K)  \in \lbrace 1,\ldots, G\rbrace^{{K}} : \forall \ g = 1,\ldots,G \quad \exists \ {k}
\textnormal{ verifying } {u}_{{k}}=g \bigr \rbrace $. In this work, we focus on the Gaussian case,  proposing different similarity criteria based on the parsimonious levels that arise from the spectral decomposition of a covariance matrix.\\

Multivariate procedures based on parsimonious decompositions assume that the theoretical covariance matrices $\Sigma_1,\ldots,\Sigma_{{K}}$ jointly verify one level $\mathscr M$ out of the fourteen in Table \ref{tabla:parsimoniosos}. To elaborate on this idea, we include now some useful notation. In a parsimonious model $\mathscr M$, we write  $(\Sigma_1,\ldots,\Sigma_{{K}}) \in \mathscr M$ if these matrices share some common parameters $C$, and they have variable parameters $\pmb V = (V_1,\ldots,V_{{K}})$ (specified in the model $\mathscr M$). We will denote by $\Sigma(V_{{k}},C)$ the covariance matrix with the size, shape and orientation parameters associated to $(V_{{k}},C)$. Therefore, under the parsimonious level $\mathscr M$, we are assuming that
\begin{equation*}
    \Sigma_{{k}} = \Sigma(V_{{k}},C) \quad \quad {k}=1,\ldots,{K} \ .
\end{equation*}
If the $n_{{k}}$ observations of group ${k}$ are independent and arise from a distribution $N(\mu_{{k}},\Sigma_{{k}})$, then $n_{{k}}S_{{k}}$ follows a ${d}$-dimensional Wishart distribution with parameters $n_{{k}},\Sigma_{{k}}$. Therefore, given the level of parsimony $\mathscr M$, it is natural to consider the maximized log-likelihood under the level $\mathscr M$ as a similarity criterion for the covariance matrices. This allows us to measure their resemblance in the features associated to the common part of the decomposition in the theoretical model. Thus, the similarity criterion for the parsimonious level $\mathscr M$ is
\begin{align*}
    f_{\mathscr M}\Bigl (\bigl \lbrace (S_{{k}},n_{{k}}),{k}=1,\ldots,r& \bigr\rbrace\Bigr) = \nonumber \\
    \underset{V_1,\ldots,V_r, C }{\max }  \sum_{{k}=1}^r \hspace{0.2cm} \log\Bigl(W_d\bigl(&n_{{k}}S_{{k}}\vert n_{{k}},\Sigma(V_{{k}},C)\bigr)\Bigr) .
\end{align*}
Consequently, given a level of parsimony $\mathscr M$, the covariance matrix classification problem in $G$ classes  consists in solving the equation
\begin{align}\label{eq:classproblem}
         {\hat{{\pmb u}}} = \underset{{\pmb u} \in \mathscr H}{\textnormal{argmax}}& \ \sum_{g=1}^G f_{\mathscr M}\Bigl(  \bigl\lbrace (S_{{k}},n_{{k}}) : {u}_{{k}}= g \bigr\rbrace \Bigr)\nonumber\\
=\underset{{\pmb u} \in \mathscr H}{\textnormal{argmax}} & \Biggl(\underset{V_1,\ldots,V_{{K}},
         C_1,\ldots,C_G  }{\max}\ \sum_{g=1}^G   \nonumber \\ 
         \sum_{{k}: {u}_{{k}}=g} \log&\Bigl(W_d\bigl(n_{{k}}S_{{k}}\vert n_{{k}},\Sigma(V_{{k}},C_g)\bigr)\Bigr)\Biggr) .
\end{align}

In order to avoid the combinatorial problem of maximizing within $\mathscr H$, denoting the variable parameters by $\pmb V=(V_1,\ldots,V_{{K}})$ and the common parameters by $\pmb C=(C_1,\ldots,C_G)$, we focus on the problem of maximizing
\begin{align*}\label{eq:classproblem2}
      W({\pmb{u}},\pmb{ V}, \pmb{C}) =&\nonumber \\
      \sum_{g=1}^G   \hspace{0.1 cm}\sum_{{k}:{u}_{{k}}=g}&\hspace{0.1cm}\log\Biggl(W_d\Bigl(n_{{k}} S_{{k}}\vert n_{{k}},\Sigma(V_{{k}},C_g)\Bigr)\Biggr) \ ,
\end{align*}
since the value ${{{\pmb u}}}$ maximizing this function agrees with the optimal ${\hat{{\pmb u}}}$ in (\ref{eq:classproblem}). This problem will be referred to as \textbf{Classification $\pmb G$-}$\pmb{\mathscr M}$. From the expression of the ${d}$-dimensional Wishart density, we can see that maximizing $W$ is equivalent to minimizing with respect to the same parameters the function

\begin{equation*}\label{eq:traza}
\sum_{g=1}^G  \sum_{{k}:{u}_{{k}}=g} n_{{k}} \Biggl( \log\Bigl(\vert\Sigma(V_{{k}},C_g)\bigr\vert+\operatorname{tr}\Bigl(\Sigma(V_{{k}},C_g)^{-1}S_{{k}}\Bigr)\Biggr) .
\end{equation*}
{Maximization} can be achieved through a simple modification of the CEM algorithm (Classification Expectation Maximization, introduced in \cite{CEM}), for any of the fourteen parsimonious levels. A sketch of the algorithm is presented here:\\ 

\noindent
\textbf{Classification} $\pmb G$-$\pmb{\mathscr M:}$ Starting from an initial estimation $\pmb{C^0}=(C_1^0,\ldots,C_G^0)$ of the common parameters, which may be taken as the parameters of $G$ different matrices $S_{{k}}$ randomly chosen between $S_1,\ldots,S_{{K}}$, the $m^{th}$ iteration consists of the following steps:

\begin{itemize}
    \item \textbf{{u}-V step}: Given the common parameters $\pmb{C^m}=(C_1^m,\ldots,C_G^m)$, we maximize with respect to the partition ${\pmb u}$ and the variable parameters $\pmb{V}$. For each ${k} = 1,\ldots,{K}$, we compute 
    \begin{align*}
    \tilde V_{{k},g} = \quad \underset{V}{\operatorname{argmax }}\ W_d\Bigl(n_{{k}} S_{{k}} \big\vert n_{{k}},\Sigma(V,C_g)\Bigr) \quad \nonumber \end{align*} 
    for $1\leq g\leq G$, and we define:
    \begin{align*}
        {u}_{{k}}^{m+1} = \underset{g\in\lbrace 1,\ldots,G\rbrace} {\operatorname{argmax }}  \ W_d\Bigl(n_{{k}} S_{{k}}  \big\vert n_{{k}}, \Sigma(\tilde V_{{k},g},C_g)\Bigr) .
    \end{align*}

    \item \textbf{V-C step:} Given the partition $\pmb{{u}^{m+1}}$, we compute the values $(\pmb{V^{m+1}},\pmb{C^{m+1}})$ maximizing $W(\pmb{{u}^{m+1}},\pmb{ V}, \pmb{C})$. The maximization can be done individually for each of the groups created, by maximizing for each $g=1,\ldots,G$ the function
    \begin{align*} 
 (\lbrace &V_{{k}} \rbrace_{k:{u_k}=g}, C_g) \longmapsto \nonumber\\
&\sum_{k:{u_k}=g}\hspace{0.1cm}\log\Bigl(W_d\bigl(n_k S_k\vert n_k,\Sigma(V_k,C_g)\bigr)\Bigr) \ ,
\end{align*}

The maximization for each of the 14 parsimonious levels can be done, for instance, with the techniques in Celeux and Govaert \citep{Parsimonious}. The methodology proposed therein for common orientation models uses modifications of the Flury algorithm \citep{FG}. However, for these models we will use the algorithms subsequently developed by \cite{McNicholas2,McNicholas1}, often implemented the software available for parsimonious model fitting, which allow more efficient estimation of the common orientation parameters. 
\end{itemize}

\noindent
For each of the fourteen parsimonious models, the variable parameters in the solution $\pmb{\hat V}$ may be computed as a function of the parameters $({\pmb{\hat{u}}},\pmb{\hat C})$, the sample covariance matrices $S_1,\ldots,S_{{K}}$ and the sample lengths $n_1,\ldots,n_{{K}}$. Therefore, the function $W$ could be written as $W(\pmb{{u}},\pmb{C})$, and the maximization could be seen as a particular case of the coordinate descent algorithm explained in \cite{coordDescent}. \\

As it was already noted, we focus on the development of the algorithm only for two particular (the most interesting) parsimonious levels. First of all, we are going to keep models flexible enough to enable the solution of (\ref{eq:intr2}), when taking $G={K}$ (no grouping is assumed), to coincide with the unrestricted solution, $\hat \Sigma_{{k}}=S_{{k}}$. The first six models do not verify this condition. For the last eight models, the numbers of parameters are 
\begin{equation*}\delta_{\textnormal{VOL}}\cdot1 + \delta_{\textnormal{SHAPE}}\cdot  (d-1) + \delta_{\textnormal{ORIENT}} \cdot \frac{d(d-1)}{2} \ ,\end{equation*}
where $\delta_{\textnormal{VOL}}, \delta_{\textnormal{SHAPE}}$ and $\delta_{\textnormal{ORIENT}}$ take the value 1 if the given parameter is assumed to be common, and ${{K}}$ if it is assumed to be variable between groups. When $d$ and ${{K}}$ are large, the main source of variation in the number of parameters is related to considering common or variable orientation, followed by considering common or variable shape. For example, if $d=9,k=6$, the number of parameters related to each constraint are detailed in Table \ref{tabla:numeroPar}.\\

\begin{table}[ht]
  \caption{Number of parameters associated with each feature when $k=6,d=9$.}
    \centering
    \medskip
    \begin{tabular}{cccc}
    \hline
      &  \hspace{0.1cm}Size\hspace{0.1cm} & \hspace{0.1cm}Shape\hspace{0.1cm} & \hspace{0.1cm}Orientation\hspace{0.1cm} \\
      \hline
       Common & 1 & 8 & 36 \\
       Variable & 6 & 48 & 216 \\
          \hline   
    \end{tabular}
    \label{tabla:numeroPar}
\end{table}

Our primary motivation is exemplified through Table \ref{tabla:numeroPar}: to raise alternatives for the models with variable orientation. For that, we  look for models with orientation varying in $G$ classes, with $1\leq G \leq {{K}}$. We  consider the case where size and shape are variable across all groups ($G$ different Common Principal Components, G-CPC) and also the case where shape parameters are additionally common within each of the $G$ classes (proportionality to $G$ different matrices, G-PROP). Apart from the parameter reduction, these models can provide an easier interpretation of the variables involved in the problem, which is often a hard task in multidimensional problems with several groups. We keep the size variable, since it does not cause a major increase in the number of parameters, and it is easy to interpret. Therefore, the models we are considering are:

\begin{itemize}
    \item  $\textbf{Classification G-CPC}$: We are looking for $G$ orthogonal matrices $\pmb{\beta}=(\beta_1,\ldots,\beta_G)$ and a vector of indexes ${\pmb{u}}=({u}_1,\ldots,{u}_{{K}}) \in \mathscr H$ such that
    \begin{equation*}
    \Sigma_{{k}} = \gamma_{{k}} \beta_{{u}_{{k}}} \Lambda_{{k}} \beta_{{u}_{{k}}}^T \quad {k}=1,\ldots,{K}\ ,
    \end{equation*}
    where $\pmb{\gamma}=(\gamma_1,\ldots,\gamma_{{K}})$ and $\pmb{\Lambda}=(\Lambda_1,\ldots,\Lambda_{{K}})$ are the variable size and shape parameters. The number of parameters is $ {{K}}+  {{K}}(d-1) + G d(d-1)/2$. In the situation of Table \ref{tabla:numeroPar}, taking $G=2$ the number of parameters  is 126, while allowing for variable orientation it is {270}. To solve (\ref{eq:classproblem}), we have to find a vector of indexes ${\pmb{\hat{u}}}$, $G$ orthogonal matrices $\pmb{\hat\beta}$ and variable parameters $\pmb{\hat\gamma}$ and $\pmb{\hat\Lambda}$ minimizing
   \begin{align}
  ({\pmb{u}},&\pmb{\Lambda},\pmb{\gamma}, \pmb{\beta}) \longmapsto \nonumber \\
  \label{eq:CPC1}
  &\sum_{g=1}^G  \sum_{{k}:{u}_{{k}}=g} n_{{k}} \left( d\log\bigl(\gamma_{{k}}\bigr)+  \frac{1}{\gamma_{{k}}}\operatorname{tr}\left(\Lambda_{{k}}^{-1}\beta_g^TS_{{k}}\beta_g\right)\right) .
   \end{align}

    \item $\textbf{Classification G-PROP}$: We are looking for $G$ orthogonal matrices $\pmb{\beta}=(\beta_1,\ldots,\beta_G)$, $G$ shape matrices $\pmb{\Lambda}=(\Lambda_1,\ldots,\Lambda_G)$ and ${\pmb u}=({u}_1,\ldots,{u}_{{K}})\in \mathscr H$ such that
    \begin{equation*}\Sigma_{{k}} = \gamma_{{k}} \beta_{{u}_{{k}}} \Lambda_{{u}_{{k}}} \beta_{{u}_{{k}}}^T \quad {k}=1,\ldots, {{K}}\ ,\end{equation*}
    where $\pmb{\gamma}=(\gamma_1,\ldots,\gamma_{{K}})$ are the variable size parameters. The number of parameters is $ {{K}}+ G (d-1) + G d(d-1)/2$. In the situation of Table \ref{tabla:numeroPar}, the number of parameters if we take $G=2$ is 94. To solve (\ref{eq:classproblem}), we have to find a vector of indexes ${\pmb{\hat{u}}}$,  $G$ orthogonal matrices $\pmb{\hat\beta}$, $G$ shape matrices $\pmb{\hat \Lambda}$ and the variable size parameters $\pmb{\hat\gamma}$ minimizing
   \begin{align}\label{eq:PROP1}
  &({\pmb{u}},\pmb{\Lambda},\pmb{\gamma}, \pmb{\beta}) \longmapsto \nonumber\\
  &\sum_{g=1}^G  \sum_{{k}:{u}_{{k}}=g} n_{{k}} \left( d\log\bigl(\gamma_{{k}}\bigr)+ \frac{1}{\gamma_{{k}}}\operatorname{tr}\left(\Lambda_g^{-1}\beta_g^TS_{{k}}\beta_g\right)\right)  .
   \end{align}
\end{itemize}

\begin{figure*}[ht] 
\centering
   \includegraphics[scale=0.75]{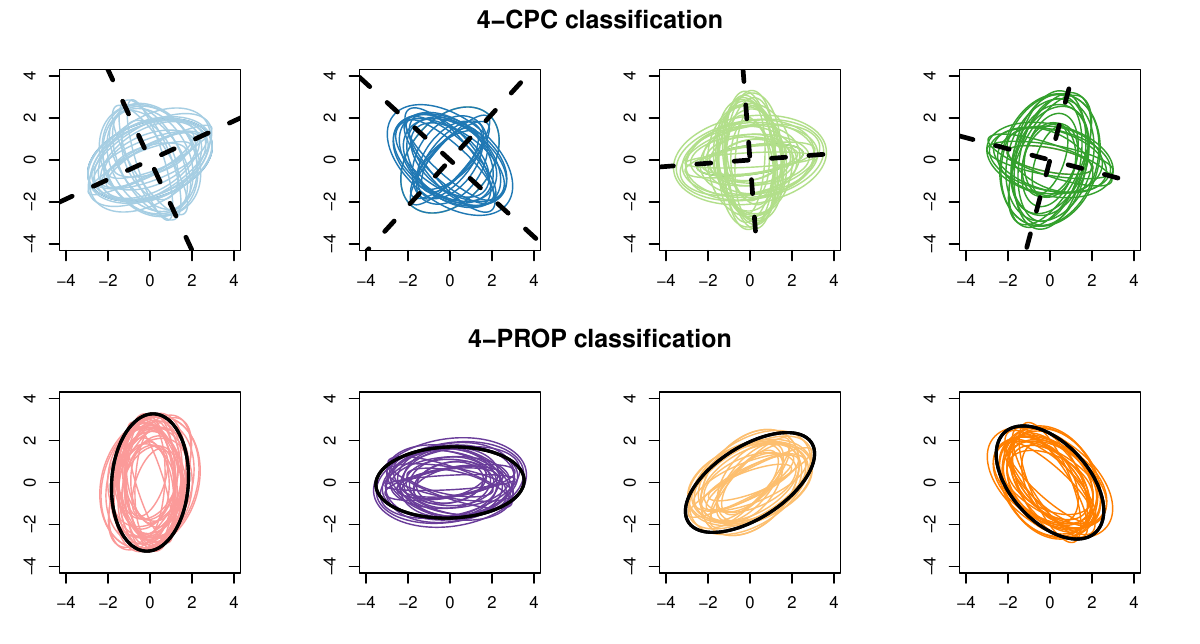}
   \caption{Classification of $S_1,\ldots,S_{100}$, represented by their 95\% confidence ellipses. The first row shows the classes and axes estimations given by the 4-CPC model, and the second row shows the classes and proportional matrix estimations given by the 4-PROP model.}
   \label{fig:clas}
 \end{figure*}

\noindent
Explicit algorithms for finding the minimum of (\ref{eq:CPC1}) and (\ref{eq:PROP1}) are given in Section \ref{alg:clas} in the Appendix. The results given by both algorithms
are illustrated in the following example, where we have randomly created 100 covariance matrices $\Sigma_1,\ldots,\Sigma_{100}$ according:
\begin{equation*}
\Sigma_{{k}} = X\Bigl(\operatorname{U}(\alpha) \operatorname{Diag}(1,Y) \operatorname{U}(\alpha)^T\Bigr)\quad  {k}=1,\ldots,100 \ ,
\end{equation*}
where $\operatorname{U}(\alpha)$ represents the rotation of angle $\alpha$, $\operatorname{Diag}(1,Y)$ is the diagonal matrix with entries $1,Y$, and $X,Y,\alpha$ are uniformly distributed random variables with distributions:
\begin{equation*}X\sim U\bigl(0.5,2\bigr)\ , \  Y\sim U\bigl(0,0.5\bigr) \ ,\ \alpha\sim U\bigl(0,\pi\bigr) \ .\end{equation*}
 
For each ${k}=1,\ldots,100$, we have taken $S_{{k}}$ as the sample covariance matrix computed from 200 independent observations from a distribution $N(0,\Sigma_{{k}})$, and we have applied 4-CPC and 4-PROP to obtain different classifications of $S_1,\ldots,S_{100}$. The partitions obtained by both methods allow us to classify the covariance matrices according to both criteria. Figure \ref{fig:clas} shows the 95\% confident ellipses representing the sample covariance matrices associated to each class (coloured lines) together with the estimations of the common axes or the common proportional matrix within each class (black lines).

\section{Gaussian Mixture Models} \label{section:GMM}
In a Gaussian Mixture Model (GMM), data are assumed to be generated by a random vector with probability density function:

\begin{equation*}f(y)= \sum_{{k}=1}^{{K}} \pi_{{k}} \phi(y\vert\mu_{{k}},\Sigma_{{k}})\ ,\end{equation*}
where $0\leq \pi_{{k}} \leq 1, \ \sum_{{k}=1}^{{K}} \pi_{{k}}=1$. The idea of introducing covariance matrix restrictions given by parsimonious decomposition in the estimation of GMMs has become a common tool for statisticians, and methods are implemented in the software $R$ in many packages. In this paper we use for the comparison the results given by the package \textit{mclust} \citep{mclust1,mclust5}, although there exists many others widely known ($Rmixmod$: \cite{Rmixmod}; $mixtools$: \cite{mixtools}). The aim of this section is to explore how we can fit GMMs in different contexts with the intermediate parsimonious models explained in Section \ref{section:clas}, allowing the common part of the covariance matrices in the decomposition to vary between $G$ classes. That is, with the same notation as in Section \ref{section:clas}, we want to study GMMs with density function
\begin{equation}\label{eq:mixtG}
    f(y)= \sum_{g=1}^G\sum_{{k}:{u}_{{k}}=g} \pi_{{k}} \phi\Bigl(y\big \vert\mu_{{k}},\Sigma(V_{{k}},C_g)\Bigr) \ ,
\end{equation}
where ${\pmb u}=({u}_1,\ldots,{u}_{{K}})\in\mathscr H$ is a fixed vector of indexes, $\pmb V = (V_1,\ldots,V_{{K}})$ are the variable parameters, $\pmb{C}=( C_1,\ldots, C_G)$ are the common parameters among classes and $\Sigma(V_{{k}},C_g)$ is the covariance matrix with the parameters given by $(V_{{k}},C_g)$. The following subsections exploit the potential of these particular GMMs for cluster analysis and discriminant analysis. A more general situation where only part of the labels are known could also be considered, following the same line as in \cite{unlabeled}, but it will not be discussed in this work. \\

As already noted in the Introduction, the criterion we are going to use for model selection between all the estimated models is BIC (Bayesian Information Criterion), choosing the model with a higher value of the BIC approximation given by
\begin{equation*}\textnormal{BIC} = 2\cdot \textnormal{loglikelihood} - \log(N)  \cdot p \ ,\end{equation*}
where $N$ is the number of observations and $p$ is the number of independent parameters to be estimated in the model. This criterion is used for the comparison of the intermediate models G-CPC and G-PROP with the fourteen parsimonious models estimated in the software $R$ with the functions in the \textit{mclust} package. In addition, within the framework of discriminant analysis, the quality of the classification given by the best models, in terms of BIC, is also compared using cross validation techniques.

\subsection{Model-Based Clustering} \label{sec:cluster}

Given $y_1,\ldots,y_N$ independent observations of a ${d}$-dimensional random vector, clustering methods based on fitting a GMM with $  {{K}}$ groups seek to maximize the log-likelihood function (\ref{eq:loglik}). From the fourteen possible restrictions considered in \cite{Parsimonious}, we can compute fourteen different maximum likelihood solutions in which size, shape and orientation are common or not between the ${{K}}$ covariance matrices. For a particular level $\mathscr M$ in Table \ref{tabla:parsimoniosos}, the fitting requires the maximization of the the log-likelihood
\begin{align*}
    L\Bigl(\pmb{\pi},\pmb{\mu} ,\pmb{V}&,C\Big\vert y_1,\ldots,y_N\Bigr) = \\
    &\sum_{{{i}}=1}^N\log\Biggl(\sum_{{k}=1}^{{K}} \pi_{{k}} \phi\Bigl(y_{{i}}\big\vert\mu_{{k}},\Sigma (V_{{k}},C)\Bigr)\Biggr) \ ,
\end{align*}
where $\pmb{\pi}=(\pi_1,\ldots,\pi_{{K}})$ are the weights, with $0\leq\pi_{{k}}\leq 1,$ $ \sum_{{k}=1}^{{K}} \pi_{{k}}=1$, $\pmb{\mu}=(\mu_1,\ldots,\mu_{{K}})$ the means,  $\pmb{V}=(V_1,\ldots,V_{{K}})$ the variable parameters and $C$ the common parameters. Estimation under the parsimonious restriction is performed via the EM algorithm. In the GMM context, we can see the complete data as pairs $(y_{{i}},z_{{i}})$, where $z_{{i}}$ is an unobserved random vector such that $z_{{i},{{k}}}=1$ if the observation $y_{{i}}$ comes from distribution ${k}$, and $z_{{i},{{k}}}=0$ otherwise. \\

With the ideas of Section \ref{section:clas}, we are going to fit Gaussian Mixture Models with parsimonious restrictions, but allowing the common parameters to vary between different classes. Assuming a parsimonious level of decomposition $\mathscr M$ and a number $G\in \lbrace 1,\ldots, {{K}}\rbrace$ of classes, we are supposing that our data are independent observations from a distribution with density function (\ref{eq:mixtG}). The log-likelihood function given a fixed vector of indexes ${\pmb u}$ is
\begin{align*} \label{eq:verSinP}
    L_{{\pmb u}}\Bigl(\pmb{\pi},\pmb{\mu}&,\pmb{V},\pmb{C}\Big\vert y_1,\ldots,y_N\Bigr) = \\ &\sum_{{{i}}=1}^N\log\Biggl(\sum_{g=1}^G\sum_{{k}:{u}_{{k}}=g} \pi_{{k}} \phi\Bigl(y_{{i}}\big\vert\mu_{{k}},\Sigma(V_{{k}},C_g)\Bigr)\Biggr)  .
\end{align*}

For each ${\pmb u}\in\mathscr H$, we can fit a model. In order to choose the best value for the vector of indexes ${\pmb u}$, we should compare the BIC values given by the different models estimated. As the number of parameters is the same, the best value for ${\pmb u}$ can be obtained by taking 
\begin{equation*}\label{eq:clusProblem}
    \pmb{{\hat{{\pmb u}}}} = \underset{{\pmb{u}} \in \mathscr H}{\textnormal{argmax}}  \Bigl[ \underset{\pmb{\pi},\pmb{\mu},\pmb{V},\pmb{C}}{\textnormal{max}}  \quad L_{{\pmb{u}}}\Bigl(\pmb{\pi},\pmb{\mu},\pmb{V},\pmb{C}\Big\vert y_1,\ldots,y_N\Bigr)  \Bigr] \ .
\end{equation*}

In order to avoid the combinatorial problem of maximizing within $\mathscr H$, we can take ${\pmb u}$ as if it were a parameter, and we are going to focus on the problem of maximizing
\begin{align} \label{eq:verG}   L\Bigl(\pmb{\pi},&\pmb{\mu},{\pmb{u}},\pmb{V},\pmb{C}\Big\vert y_1,\ldots,y_N\Bigr) = \nonumber\\
&\sum_{{{i}}=1}^N\log\Biggl(\sum_{g=1}^G\sum_{{k}:{u}_{{k}}=g} \pi_{{k}} \phi\Bigl(y_{{i}}\big\vert\mu_{{k}},\Sigma(V_{{k}},C_g)\Bigr)\Biggr)  ,
\end{align}
that will be referred to as \textbf{Clustering G-}$\pmb{\mathscr M}$. Therefore, given the unobserved variables $z_{{i},{{k}}}$, for ${k}=1,\ldots,{K}$ and ${{i}}=1,\ldots,N$, the complete log-likelihood is

\begin{align} \label{eq:verCompleta}
&CL\Bigl(\pmb{\pi},\pmb{\mu} ,{\pmb u}, \pmb{V},\pmb C\Big\vert 
y_1,\ldots,y_N,z_{1,1},\ldots, z_{{N,K}}\Bigr) = \nonumber \\
&\sum_{{{i}}=1}^N \left[\sum_{g=1}^G\sum_{{k}:{u}_{{k}}=g} z_{{i},{{k}}} \log\Biggl( \pi_{{k}} \phi\Bigl(y_{{i}}\big\vert\mu_{{k}},\Sigma(V_{{k}},C_g)\Bigr)\Biggr)\right] \ .
\end{align}

The proposal of this section is to fit this model given a parsimonious level $\mathscr M$ and fixed values of  ${{K}}$ and $G \in \lbrace 1,\ldots, {{K}}\rbrace$, introducing also constraints to avoid the unboundedness of the log-likelihood function (\ref{eq:verG}). For this purpose, we introduce the determinant and shape constraints studied in \cite{detShape}. For ${k}=1,\ldots,{K}$, denote by $(\lambda_{{k},1},\ldots,\lambda_{{k},d})$ the diagonal elements of the shape matrix $\Lambda_{{k}}$ (which may be the same within classes). We impose ${{K}}$ constraints controlling the shape of each group, in order to avoid solutions that are almost contained in a subspace of lower dimension, and a size constraint in order to avoid the presence of very small clusters. Given $c_{sh},c_{vol}\geq 1$, we impose:
\begin{equation}\label{eq:shape}
    \frac{\underset{l=1,\ldots,d}{\operatorname{max}}\lambda_{{k},l}}{\underset{l=1,\ldots,d}{\operatorname{min}}\lambda_{{k},l}}\leq c_{sh}, \ {k}=1,\ldots,{K},\quad  
        \frac{\underset{{k}=1,\ldots,{K}}{\operatorname{max}}\gamma_{{k}}}{\underset{{k}=1,\ldots,{K}}{\operatorname{min}}\gamma_{{k}}}\leq c_{vol} \ ,
\end{equation}

\begin{remark}\label{remark}
With these restrictions, the theoretical problem of maximizing (\ref{eq:verG}) is well defined. If $Y$ is a random vector following a distribution $\mathbb P$, the problem consists in maximizing
\begin{align} \label{eq:teor}
&\operatorname{E}\Biggl[\log\Biggl(\sum_{g=1}^G\sum_{{k}:{u}_{{k}}=g} \pi_{{k}} \phi\Bigl( Y \thinspace  \big\vert\mu_{{k}},\Sigma\bigl(V_{{k}},C_g\bigr)\Bigr)\Biggr)\Biggr] = \nonumber \\
&\bigintssss\log\Biggl(\sum_{g=1}^G\sum_{{k}:{u}_{{k}}=g} \pi_{{k}} \phi\Bigl( y \thinspace  \big\vert\mu_{{k}},\Sigma\bigl(V_{{k}},C_g\bigr)\Bigr)\Biggr) \operatorname{d\mathbb P}(y)
\end{align}
with respect to $\pmb{\pi},\pmb{\mu},{\pmb{u}} ,\pmb{V},\pmb{C}$, defined as above, and verifying (\ref{eq:shape}). If ${\mathbb P} _N$ stands for the empirical measure ${\mathbb P} _N=(1/ N) \sum_{{i}=1}^N \delta_{\left\{y_{{i}}\right\}}$, by replacing ${\mathbb P} $ by ${\mathbb P} _N$, we recover the original sample problem of maximizing (\ref{eq:verG}) under the determinant and shape constraints (\ref{eq:shape}). This approach guarantees that the objective function is bounded, allowing results to be stated in terms of existence and consistence of the solutions (see Section  \ref{app:teor} in the Appendix).
\end{remark}

Now, we are going to give a sketch of the EM algorithm used for the estimation of these intermediate parsimonious clustering models, for each of the fourteen levels.\\

 \textbf{Clustering G-}$\pmb{\mathscr M: }$ Starting from an initial solution of the parameters $\pmb{\pi^{0}},\pmb{\mu^{0}},\pmb{{u}^{0}}$, $\pmb{V^{0}},\pmb{C^{0}}$, we have to repeat the following steps until convergence:

\begin{itemize}
    \item \textbf{E step}: Given the current values of the parameters $\pmb{\pi^{m}},\pmb{\mu^{m}},\pmb{{u}^{m}}$, $\pmb{V^{m}},\pmb{C^{m}}$, we compute the posterior probabilities
    \begin{equation}\label{eq:pasoE}
        z_{{i},{{k}}} = \frac{\pi_{{k}}^m \phi\Bigl(y_{{i}}\vert\mu_{{k}}^m,\Sigma\bigl(V_{{k}}^m,C_{{u}_{{k}}}^m\bigr)\Bigr)}{\sum_{l=1}^{{K}} \pi_l^m \phi\Bigl(y_{{i}}\vert\mu_l^m,\Sigma\bigl(V_l^m,C_{{u}_l}^m\bigr)\Bigr)} \end{equation}
        for ${k}=1,\ldots,{K}, \ {{i}}=1,\ldots,N$.
    \item \label{eq:pasoM}\textbf{M step}: In this step, we have to maximize (\ref{eq:verCompleta}) given the expected values $\lbrace z_{{i},{{k}}} \rbrace_{{i},{{k}}}$. The optimal values for $\pmb{\pi^{m+1}},\pmb{\mu^{m+1}}$ are given by (\ref{eq:optValues}). With these optimal values, if we denote $S_{{k}}= (1/n_{{k}})\sum_{{{i}}=1}^N z_{{i},{{k}}} (y_{{i}}-\mu_{{k}}^{m+1})(y_{{i}}-\mu_{{k}}^{m+1})^T$, then we have to find the values $\pmb{{u}^{m+1}}$, $\pmb{V^{m+1}},\pmb{C^{m+1}}$ verifying the determinant and shape constraints (\ref{eq:shape}) maximizing
    \begin{align*}
       ({\pmb u},  &\pmb{V},\pmb C)\longmapsto \\
       &CL\Bigl(\pmb{\pi^{{m+1}}},\pmb{\mu^{{m+1}}} ,{\pmb u},\pmb{V},\pmb C\Big\vert \{y_i\}_{i},\{z_{i,k}\}_{i,k}\Bigr) \ . 
    \end{align*}
    If we remove the determinant and shape constraints, the solution of this maximization coincides with the classification problem presented in Section \ref{section:clas} for the computed values of $n_1,\ldots,n_{{K}}$ and $S_1,\ldots,S_{{K}}$. A simple modification of that algorithm, computing on each step the optimal size and shape constrained parameters (instead of the unconstrained version) with the \textit{optimal truncation} algorithm presented in \cite{detShape} allows the maximization to be completed. Determinant and shape constraints can be incorporated in the algorithms together with the parsimonious constraints following the lines developed in \cite{ultimo}.
\end{itemize}

\begin{figure*}[ht] 
\centering
   \includegraphics[scale=0.7]{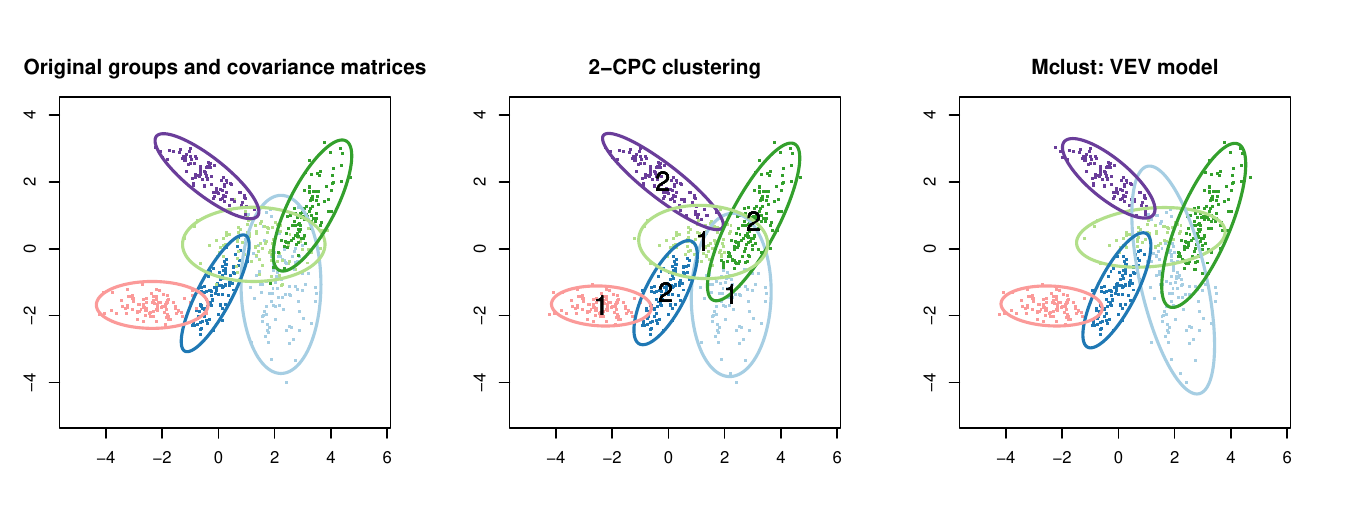}
   \caption{From left to right: 1. Theoretical Gaussian distributions and observations simulated from each distribution. 2. Solution estimated by clustering through 2-CPC model. 3. Best clustering solution estimated by \textit{mclust} in terms of BIC.}
   \label{fig:2-CPC}
 \end{figure*}

 As already noted in Section \ref{section:clas}, we keep only the clustering models G-CPC and G-PROP, the most interesting in terms of parameter reduction and interpretability. For these models, explicit algorithms are explained in Section \ref{alg:cluster} in the Appendix. Now, we are going to illustrate the results of the algorithms in two simulation experiments:

\begin{itemize}
    \item \textbf{Clustering G-CPC}: In this example, we simulate $n=100$ observations from each of 6 Gaussian distributions, with means $\mu_1,\ldots,\mu_6$ and covariance matrices verifying
    \begin{align*}
       \Sigma_{{k}} = &{\gamma}_{{k}}\beta_1\Lambda_{{k}} \beta_1^T, \quad {k}=1,2,3, \\ \Sigma_{{k}} = &{\gamma}_{{k}}\beta_2\Lambda_{{k}} \beta_2^T ,\quad {k}=4,5,6 \ . 
    \end{align*}
    
In Figure \ref{fig:2-CPC}, we can see in the first plot the 95 \% confidence ellipses of the six theoretical Gaussian distributions together with the 100 independent observations simulated from these distributions. The second plot represents the clusters created by the maximum likelihood solution for the 2-CPC model, taking $c_{sh}=c_{vol}=100$. The numbers labeling the ellipses represent the class of covariance matrices sharing the orientation. Finally, the third plot represents the best solution estimated by \textit{mclust} for ${K}=6$, corresponding to the parsimonious model VEV, with equal shape and variable size and orientation. The BIC value in the 2-CPC model (31 d.f.) is -3937.08, whereas the best model VEV  (30 d.f.) estimated with \textit{mclust} has BIC value -3960.07. Therefore, the GMM estimated with the 2-CPC restriction has higher BIC than all the parsimonious models. Finally, the number of observations assigned to different clusters from the original ones is 82 for the 2-CPC model and 91 for the VEV model. \\

  \item \textbf{Clustering G-PROP}: In this example, we simulate $n=100$ observations from each of 6 Gaussian distributions, with means $\mu_1,\ldots,\mu_6$ and covariance matrices verifying:
    \begin{align*}
        \Sigma_{{k}} = &{\gamma}_{{k}} A_1, \quad {k}=1,2,3, \\ \Sigma_{{k}} =& {\gamma}_{{k}} A_2 ,\quad {k}=4,5,6 \ .
    \end{align*}
    Figure \ref{fig:2-PROP} is analogous to  Figure \ref{fig:2-CPC}, but in the proportionality case. The BIC value for the 2-PROP  model (27 d.f.) with $c_{sh}=c_{vol}=100$ is -3873.127, whereas the BIC value for the best model fitted by \textit{mclust} is -3919.796, which corresponds to the unrestricted model VVV (35 d.f.). Now, the number of observations wrongly assigned to the source  groups is 64 for the  2-PROP model, while it is 71 for the VVV model. 
     \begin{figure*}[ht] 
    \centering
       \includegraphics[scale=0.7]{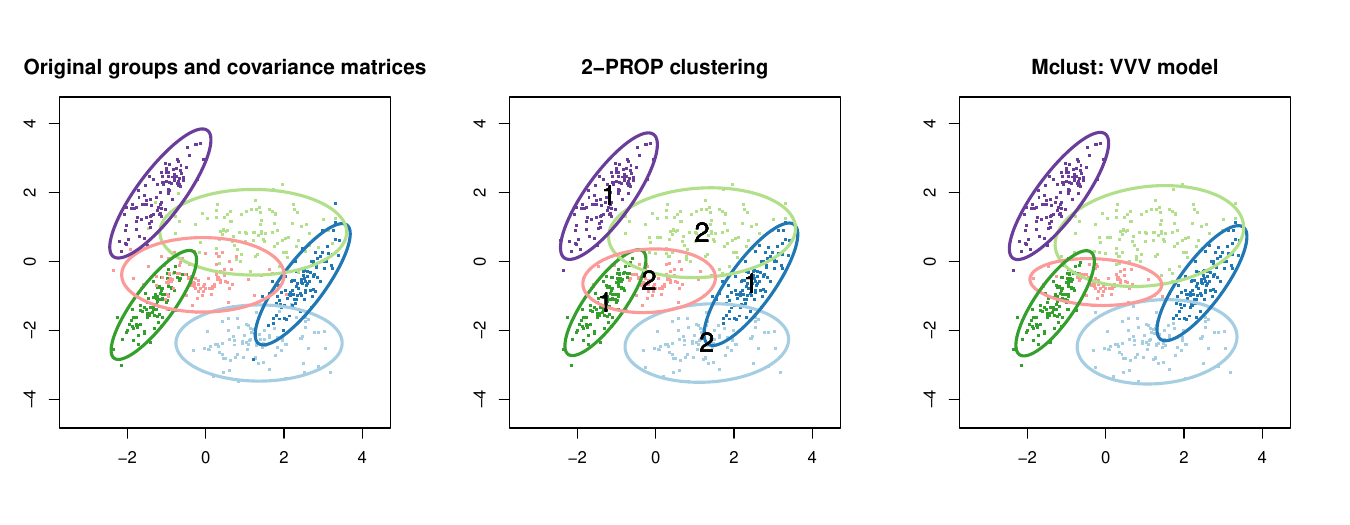}
       \caption{From left to right: 1. Theoretical Gaussian distributions and observations simulated from each distribution. 2. Solution estimated by clustering through 2-PROP model. 3. Best clustering solution estimated by \textit{mclust} in terms of BIC.}
       \label{fig:2-PROP}
     \end{figure*}
\end{itemize}

\begin{remark}
{
Note that, by imposing appropriate constraints in the clustering problem, we can significantly decrease the number of parameters while keeping a good fit of the data. {Figure \ref{fig:2-PROP} shows this effect.} However, constraints also have a clear interpretation in cluster analysis problems, since we are looking for groups that are forced to have a particular shape. Therefore, different constraints can lead to clusters with different shapes. This is what happens in Figure \ref{fig:2-CPC}, where by introducing the right constraints we have managed to make the clusters created more similar to the original ones. Of course, in the absence of prior information, it is not possible to know the appropriate constraints, and the most reasonable approach is to select a model according to a criterion that penalizes the fit with the number of parameters such as the BIC.    }
\end{remark} 

To evaluate the sensitivity of {BIC} for the detection of the true underlying model, we have used the models described in the two previous examples. Once a model and a particular {sample size $n$} (=50, 100, 200) have been chosen, the simulation planning produces a sample containing $n$ random elements generated from each $N(\mu_{{k}},\Sigma_{{k}}),\  {k}=1,\ldots,6$. We repeated every simulation plan 1000 times, comparing for every sample the BIC obtained for the underlying clustering model vs the best parsimonious model estimated by \textit{mclust}. Table \ref{proportions} includes the proportions of times in which 2-CPC or 2-PROP model improves the best \textit{mclust} model {in terms of BIC} for each value of $n$. Of course, the accuracy of the approach should depend on the dimension, the number of groups, the overlapping... However, even in the case of a large overlapping, as in the present examples, the proportions reported in Table \ref{proportions} show that moderate values of ${n}$ suffice to get very high proportions of success. {Appendix \ref{app:additionalSim} contains additional simulations supporting the suitability of BIC in this framework.}

\begin{table}[ht]
\caption{Proportions of times in which clustering 2-CPC or 2-PROP model improves the best \textit{mclust} model in terms of BIC, for each  {sample size $n$}. }

    \centering
    \begin{tabular}{cccc}
        \hline 
        \hspace{0.1cm}Example \hspace{0.1cm}& \hspace{0.1cm}n=50\hspace{0.1cm} & \hspace{0.1cm}n=100\hspace{0.1cm}&\hspace{0.1cm} n=200\hspace{0.1cm} \\
        \hline 
         2-CPC & 0.570 & 0.927 & 1.000 \\
         2-PROP& 0.933 & 0.999 & 1.000 \\
         \hline
    \end{tabular}
    \label{proportions}
\end{table}

\subsection{Discriminant Analysis}

\begin{table*}[t]
\caption{Classification results for data in Figure \ref{fig:2-CPC} for the best \textit{mclust} model and 2-CPC.}
\medskip
    \centering
    \begin{tabular}{c|cccccc}
    \hline
       model & loglik & df &  BIC & MM & LOO & CV(300,0.9) \\
      \hline
      \textit{mclust}: VVV & -1874.865 & 30 & -3941.637 & 66/600 &  71/600  & 0.1187  \\
      2-CPC &  -1874.74 &  26 &  \textbf{-3915.801} & 65/600 &  69/600  &   0.1161 \\
          \hline   
    \end{tabular}
    \label{tab:res1}
\end{table*}

\begin{table*}[ht]
\caption{Classification results for data in Figure \ref{fig:2-PROP} for the best \textit{mclust} model and 2-PROP.}
\medskip
    \centering
    \begin{tabular}{c|cccccc}
    \hline
       model & loglik & df &  BIC & MM & LOO & CV(300,0.9) \\
      \hline
      \textit{mclust}: VVV & -1852.765&  30 & -3897.439 & 62/600 &  69/600  & 0.1102  \\
      2-PROP &  -1853.056 &  22 & \textbf{-3846.845} & 64/600 &  68/600  &   0.1083 \\
          \hline   
    \end{tabular}
    \label{tab:res2}
\end{table*}

The parsimonious model introduced in \cite{discriminante2} for discriminant analysis  has been developed in conjunction with model-based clustering. The $R$ package \textit{mclust} \citep{mclust1,mclust5} also includes functions for fitting these models, denoted by EDDA (Eigenvalue Decomposition Discriminant Analysis). In this context, given a parsimonious level $\mathscr M$ and a number $G$ of classes, we can also consider fitting an intermediate model for each fixed ${\pmb u}\in \mathscr H$, by maximizing the complete log-likelihood
\begin{align} \label{eq:verCompletaP}
C&L_{{\pmb u}}\Bigl(\pmb{\pi},\pmb{\mu} , \pmb{V},\pmb C\Big\vert y_1,\ldots,y_N,z_{1,1},\ldots,z_{{N,K}}\Bigr) = \nonumber \\
&\sum_{{{i}}=1}^N \left[\sum_{g=1}^G\sum_{{k}:{u}_{{k}}=g} z_{{i},{{k}}} \log\Biggl( \pi_{{k}} \phi\Bigl(y_{{i}}\vert\mu_{{k}},\Sigma(V_{{k}},C_g)\Bigr)\Biggr)\right] \ .
\end{align}\newpage

Model comparison is done through BIC, and consequently we could try to choose ${\pmb u}$ maximizing the log-likelihood (\ref{eq:verG}). However, given that in the model fitting we are maximizing the complete log-likelihood (\ref{eq:verCompletaP}), it is not unreasonable trying to find the value of ${\pmb u}$ maximizing (\ref{eq:verCompletaP}). Proceeding in this manner, we can think of ${\pmb u}$ as a parameter, and the problem consists in maximizing (\ref{eq:verCompleta}). Model estimation is simple from model-based clustering algorithms: with a single iteration of the M step, we can compute the values of the parameters. A new set of observations can be classified computing the posterior probabilities, with the formula (\ref{eq:pasoE}) of the E step, and assigning each new observation to the group with higher posterior probability. Since the groups are known, the complete log-likelihood (\ref{eq:verCompleta}) is bounded under mild conditions, and it is not required to impose eigenvalue constraints, although it may be interesting in some examples with almost degenerated variables. To summarize the quality of the classification given by the best models (selected through BIC) in the different examples, other indicators based directly on classification errors are provided:\\

\begin{itemize}
    \item[-] \textbf{MM}: Model Misclassification, or training error. Proportion of observations misclassified by the model fitted with all observations.
    
    \item[-] \textbf{LOO}: Leave One Out error.
    
    \item[-] \textbf{CV({R},p):} Cross Validation error. Considering each observation as labeled or unlabeled with probability $p$ and $1-p$, we compute the proportion of unlabeled observations misclassified by the model fitted with the labeled observations. The indicator CV({R},p) represents the mean of the proportions obtained in {R} repetitions of the process. When several classification methods are compared, the same {R} random partitions are used to compute the values of  this indicator. 
\end{itemize}

\noindent
In the line of the previous section, only the discriminant analysis models G-CPC and G-PROP are considered.  Table \ref{tab:res1} and \ref{tab:res2} show the results of applying these models to the simulation examples of Figure \ref{fig:2-CPC}, \ref{fig:2-PROP}. In both situations, the classification obtained with our model slightly improves that given by \textit{mclust}.\\

As we did in the clustering setting, in order to evaluate the sensitivity of {BIC} for the detection of the true underlying model, simulations have been repeated 1000 times, for each {sample size $n$} (=30, 50, 100, 200). Table \ref{tab:disc} shows the proportions of times in which 2-CPC or 2-PROP model improves the best \textit{mclust} model {in terms of BIC} for each value of ${n}$.

\begin{table}[ht]
\caption{Proportions of times in which discriminant analysis 2-CPC or 2-PROP model improves the best \textit{mclust} model in terms of BIC, for each  {sample size $n$}. }
  \medskip
    \centering
    \begin{tabular}{ccccc}
        \hline 
        \hspace{0.1cm}Example\hspace{0.1cm} & \hspace{0.1cm}n=30\hspace{0.1cm} & \hspace{0.1cm}n=50\hspace{0.1cm} & \hspace{0.1cm}n=100\hspace{0.1cm}&\hspace{0.1cm} n=200\hspace{0.1cm} \\
        \hline 
         2-CPC &  0.443& 0.782 & 0.975 & 1.000 \\
         2-PROP &0.971& 1.000 & 1.000 & 1.000 \\
         \hline
    \end{tabular}
    \label{tab:disc}
\end{table}
\begin{remark}\label{remark2}
In discriminant analysis, the weights $\pmb \pi=(\pi_1,\ldots,\pi_{{K}})$ might not be considered as parameters. Model-based methods assume that observations from the ${k}^{th}$ group follow a distribution with density function $f(\cdot,\theta_{{k}})$. If $\pi_{{k}}$ is the proportion of observations of group ${k}$, the classifier minimizing the expected misclassification rate is known as Bayes classifier, and it assigns an observation $y$ to the group with higher posterior probability
\begin{equation}\label{eq:Bayes}
    P\bigl(y \in \textnormal{Group }{k}\bigr) = \frac{\pi_{{k}} f(y,\theta_{{k}})}{\sum_{l=1}^{{K}} \pi_l f(y,\theta_l)} \ .
\end{equation}
The values of $\pmb \pi,\theta_1,\ldots,\theta_{{K}}$ are usually unknown,
and the classification is performed with estimations $\pmb{\hat \pi},\hat\theta_1,\ldots,\hat\theta_{{K}}$. Whereas $\hat\theta_1,\ldots,\hat\theta_{{K}}$ are always parameters estimated from the sample, the values of $\pmb{\hat \pi}$ may be seen as part of the classification rule, if we think that they represent a characteristic of a particular sample we are classifying, or real parameters, if we assume that the observations $(z_{{i}},y_{{i}})$ arise from a GMM such that
\begin{align*}
 z_{{i}} \sim \operatorname{mult}\Bigl(1,\lbrace 1,\ldots,  {{K}}\rbrace,&\lbrace \pi_1,\ldots, \pi_{{K}} \rbrace \Bigr)\\
 y_{{i}} \big\vert z_{{{i}}} \sim f\bigl(&\cdot, \theta_{z_{{i}}}\bigr) \ , 
\end{align*}
where $mult()$ denotes the multinomial distribution, and the weights verify $0\leq\pi_{{k}}\leq  1$, $\sum_{{k}=1}^{{K}} \pi_{{k}} =1$. In accordance with \textit{mclust}, for model comparison we are not considering $\pmb{\pi}$ as parameters, although its consideration would only mean adding a constant to all BIC values computed. However, in order to define the theoretical problem, the situation where we are considering $\pmb{\pi}$ as a parameter is more interesting. If $(Z,Y)$ is a random vector following a distribution $\mathbb P$ in $\lbrace 1,\ldots, {{K}}\rbrace \times \mathbb R^d$, the theoretical problem consists in maximizing
\begin{align}\label{eq:teorDA}
&\operatorname{E}\Biggl[\sum_{g=1}^G\sum_{{k}:{u}_{{k}}=g} \operatorname{I}(Z={k}) \log\Biggl( \pi_{{k}} \phi\Bigl(Y\vert\mu_{{k}},\Sigma(V_{{k}},C_g)\Bigr)\Biggr)\Biggr] =  \nonumber\\
  &\bigintssss \hspace*{-0.1cm}\sum_{g=1}^G \hspace*{-0.2cm}\sum_{\hspace*{0.2cm}{k}:{u}_{{k}}=g}\hspace*{-0.28cm}\operatorname{I}(z={k}) \log\Biggl( \hspace*{-0.1cm}\pi_{{k}} \phi\Bigl(y\vert\mu_{{k}},\Sigma(V_{{k}},C_g)\Bigr)\Biggr)\hspace*{-0.1cm} \operatorname{d\mathbb P}(z,y)
\end{align}
with respect to the parameters $\pmb \pi, \pmb \mu, {\pmb u},\pmb V, \pmb C$. Given $N$ observations $(z_{{i}},y_{{i}}), \ {{i}}=1,\ldots,N$ of $\mathbb P$, the problem of maximizing (\ref{eq:teorDA}) agrees with the sample problem presented above the remark when taking the empirical measure $\mathbb P_N$, with the obvious relation $z_{{i},{{k}}}=\operatorname{I}(z_{{i}}={k})$. Arguments like those presented in Section \ref{app:teor} in the Appendix for the cluster analysis problem would give existence and consistency of solutions also in this setting.

\end{remark}

\section{Real Data Examples}\label{section:examples}

To illustrate the usefulness of the G-CPC and G-PROP models in both settings, we show four real data examples in which our models outperform the best parsimonious models fitted by \textit{mclust}, in terms of BIC. The two first examples are intended to illustrate the methods in simple and well-known data sets, while the latter involve greater complexity.

\subsection{Cluster Analysis: IRIS}

Here we revisit the famous \textit{Iris data set}, which consists of observations of four features (length and width of sepals and petals) of 50 samples of three species of Iris (setosa, versicolor and virginica), and is available in the base package of $R$. We apply the functions of package \textit{mclust} for model-based clustering, letting the number of clusters to search equal to 3, to obtain the best parsimonious model in terms of BIC value. Table \ref{tabla:Iris} compares this model with the models 2-CPC and 2-PROP, fitted with $c_{sh}=c_{vol}=100$. With some abuse of notation, we include in the table the Model Misclassification (MM), representing here the number of observations assigned to different clusters than the originals, after identifying the clusters created with the originals in a logical manner.

\begin{table}[ht]
  \caption{Iris data solutions for clustering with \textit{mclust}, 2-CPC and 2-PROP.}
\medskip
    \centering
    \begin{tabular}{c|cccc}
    \hline
       model & loglik & df &  BIC & MM \\
      \hline
      \textit{mclust}: VEV & -186.074  & 38 & -562.550 & 5/150 \\
      2-CPC &        -185.538 & 38 &  -561.480 &   5/150  \\
      2-PROP &       -192.177 & 35 & \textbf{-559.727} &  4/150 \\
          \hline   
    \end{tabular}
    \label{tabla:Iris}
\end{table}

 \begin{figure*}[ht] 
    \centering
       \includegraphics[scale=0.4,angle=0]{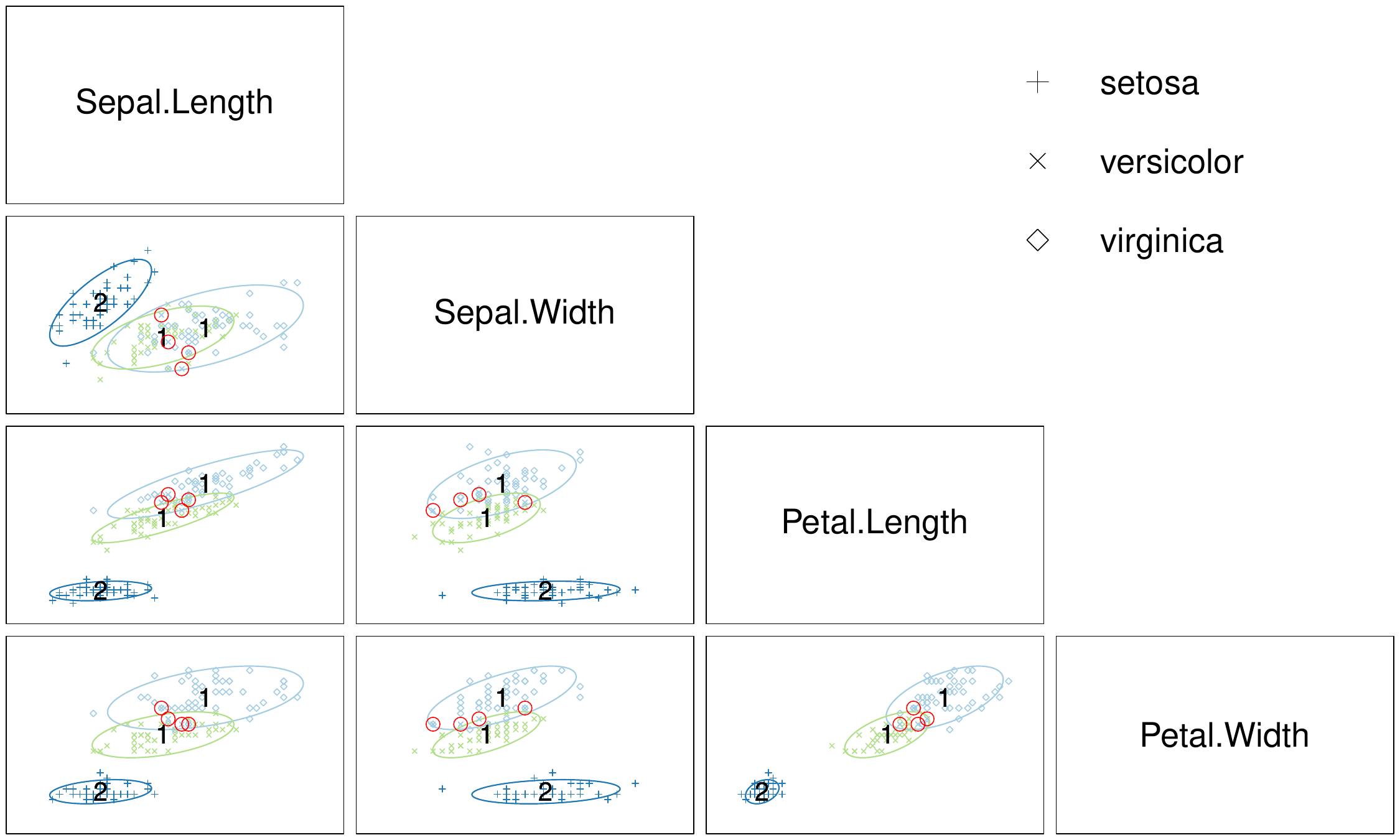}
      \caption{Clustering obtained from 2-PROP model in the Iris data set. Color represents the clusters created. The ellipses are the contours of the estimated mixture densities, grouped into the classes given by indexes in black. Point shapes represent the original groups. Observations lying on different clusters from the originals are marked with red circles.}
       \label{fig:iris}
\end{figure*}
    
From Table \ref{tabla:Iris} we can appreciate that the best clustering model in terms of BIC is the 2-PROP model. In Figure \ref{fig:iris} we can see the clusters created by this model. These clusters coincide with the real groups, except for four observations. From this example, we can also see the advantage of the intermediate models G-CPC and G-PROP in terms of interpretability. In the solution found with G-PROP the covariance matrices associated to two of the three clusters are proportional. Each cluster represents a group of individuals with similar features, which in absence of labels, we could see as a subclassification within the Iris specie. In this subclassification associated to the groups with proportional covariance matrices, both groups share not only the  principal directions, but also the same proportion of variability between the directions. In many biological studies, principal components are of great importance. When working with phenotypic variables, principal components may be interpreted as ``growing directions" (see e.g. \cite{Thorpe}). From the estimated model, we can conclude that in the Iris data, it is reasonable to think that there are three groups, two of them with similar ``growing pattern", since not only the principal components are the same, but also the shape is common. However, this biological interpretation will become even more evident in the following example.



\subsection{Discriminant Analysis: CRABS}

The data set consists of measures of 5 features over a set of 200 crabs from two species, orange and blue, and from both sexes, and it is available in the  $R$ package \textit{MASS} \citep{MASS}. For each specie and sex (labeled OF, OM, BF, BM) there are 50 observations. The variables are measures in mm of the following features: frontal lobe (FL), rear width (RW), carapace length (CL), carapace width (CW) and body depth (BD). Applying the classification function of the \textit{mclust} library, the best parsimonious model in terms of BIC is EEV. Table \ref{tabla:Crabs} shows the result for the EEV model, together with the discriminant analysis models 2-CPC and 2-PROP, with $c_{sh}=c_{vol}=100000$ (with these values, the solutions agrees with the unrestricted solutions).
\begin{table*}[ht]
 \caption{Crabs data solutions for discriminant analysis with \textit{mclust}, 2-CPC and 2-PROP.}
 \medskip
    \centering
    \begin{tabular}{c|ccccccc}
    \hline
       model & loglik & df &  BIC & MM & LOO & CV(300,0.8) & CV(300,0.95)\\
      \hline
      \textit{mclust}: EEV & -1247.693   & 65 & -2839.776   & 8/200 & 9/200 & 0.0513 &   0.0521 \\
      2-CPC &         -1271.470   & 60 &  -2860.839  & 7/200 &  9/200  & 0.0536 &  0.0514 \\
      2-PROP &    -1278.906  & 52 & \textbf{-2833.324}& 8/200&  11/200 & 0.0546  & 0.0613\\
          \hline   
    \end{tabular}
    \label{tabla:Crabs}
\end{table*}
The results show that the comparison given by BIC can differ from those obtained by cross validation techniques, partially because BIC mainly measures the fit of the data to the model. However, in the parsimonious context, model selection is usually performed via BIC, in order to avoid the very time-consuming process of evaluating every possible model with cross validation techniques.\\

Figure \ref{fig:crabs} represents the solution estimated by 2-PROP model. The solution given by this model allows for a better biological interpretation  than the  one given by the parsimonious model EEV, where orientation varies along the 4 groups, making the comparison quite complex. In the 2-PROP model, the groups of males of both species share proportional matrices, and the same is true for the females. Returning to the biological interpretation of the previous example, under the 2-PROP model, we can state that crabs of the same sex have the same ``growing pattern'', despite of being from different species. 

\subsection{Cluster Analysis: GENE EXPRESSION CANCER}

In this example, we work with the \textit{Gene expression cancer RNA-Seq Data Set}, which can be downloaded from the \href{https://archive.ics.uci.edu/ml/datasets/gene+expression+cancer+RNA-Seq}{UCI Machine Learning Repository}. This data set is part of the data collected by ``The Cancer Genome Atlas Pan-Cancer
analysis project"" \citep{cancer}. The {considered data set} consists of a random extraction of gene expressions of patients having different types of tumor: BRCA (breast carcinoma), KIRC (kidney renal
clear-cell carcinoma), COAD (colon adenocarcinoma), LUAD (lung squamous carcinoma) and PRAD (prostate adenocarcinoma). In total, the data set contains the information of 801 patients, and for each patient we have information of 20531 variables, which are the RNA sequencing values of 20531 genes. To reduce the dimensionality and to apply model-based clustering algorithms, we have removed the genes with almost zero sum of squares ($< 10^{-5}$) and applied PCA to the remaining genes. We have taken the first 14 principal components, the minimum number of components retaining more than 50 $\%$ of the total variance. Applying model-based clustering methods looking for 5 groups to this reduced data set, we have found that 3-CPC, fitted with $c_{sh}=c_{vol}=1000$, improves the BIC value obtained by the best parsimonious model estimated by \textit{mclust}. The results obtained from 3-CPC, presented in Table \ref{tab:Cancer}, significantly improve the assignment error made by $mclust$. Figure \ref{fig:cancer3cpc}  shows the projection of the solution obtained by 3-CPC onto the first six principal components computed in the preprocessing steps.  
\begin{table}[ht]
    \caption{Cancer data solutions for clustering with \textit{mclust} and 3-CPC.}
 \medskip
    \centering
    \begin{tabular}{c|cccc}
    \hline
       model & loglik & df &  BIC & MM \\
      \hline
      \textit{mclust}: VVV & -44121.24 &  599 &       -92247.32     & 64/801  \\
      3-CPC &         -44561.12 &  417 &  \textbf{-91910.25} & 6/801   \\
          \hline   
    \end{tabular}
    \label{tab:Cancer}
\end{table}

\subsection{Discriminant Analysis: ITALIAN OLIVE OIL}

\begin{table*}[ht]
    \caption{Olive oil discriminant analysis with \textit{mclust}, 2-CPC, 3-CPC and 3-PROP.}
 \medskip
    \centering
    \begin{tabular}{c|ccccccc}
    \hline
       model & loglik & df &  BIC & MM & LOO & CV(300,0.8) & CV(300,0.95)\\
      \hline
      \textit{mclust}: VVE &-20595.49 & 172  & -42283.03   & 12/572 &  20/572  & 0.0375 &   0.0363\\
      2-CPC &      -20452.64   &  200 &   -42175.11  & 10/572 & 18/572  & 0.0369  & 0.0281 \\
      3-CPC & -20332.93& 228 &  \textbf{-42113.47} & 9/572&  16/572 & 0.0365  & 0.0278\\
      3-PROP &  -20521.33 & 186 & -42223.60         & 16/172 & 27/572& 0.0464  & 0.0463 \\
          \hline   
    \end{tabular}
    \label{tab:Oil}
\end{table*}

The data set contains information about the composition in percentage of eight fatty acids (palmitic, palmitoleic, stearic, oleic, linoleic, linolenic, arachidic and eicosenoic) found in the lipid fraction of 572 Italian olive oils, and it is available in the $R$ package \textit{pdfCluster} \citep{pdfCluster}. The olive oils are labeled according to a two level classification: 9 different areas that are grouped  at the same time in three different regions. 
\begin{itemize}
    \item SOUTH: Apulia North, Calabria, Apulia South, Sicily.
    \item SARDINIA: Sardinia inland, Sardinia coast.
    \item CENTRE-NORTH: Umbria, Liguria east, Liguria west.
\end{itemize}
In this example, we have evaluated the performance of different discriminant analysis models, for the problem of classifying the olive oils between areas. The best parsimonious model fitted with \textit{mclust} is the VVE model, with variable size and shape and equal orientation. Note that  due to the dimension $d=8$, there is a significant difference in the number of parameters between models with common or variable orientation. Therefore, BIC selection will tend to choose models with common orientation, despite the fact that this hypothesis might not be very precise. This  suggests that intermediate models could be of great interest also in this example. Given that the last variable \textit{eicosenoic} is almost degenerated in some areas, we fit the models with $c_{sh}=c_{vol}=10000$, and the shape constraints are effective in some groups. We have found 3 different intermediate models improving the BIC value obtained with \textit{mclust}. Results are displayed in Table \ref{tab:Oil}.\\

The best solution found in terms of BIC is given by the 3-CPC model, which is also the solution with the best values for the other indicators. The classification of the areas in classes given in this solution is:
\begin{itemize}
    \item CLASS 1:  Umbria.
    \item CLASS 2: Apulia North, Calabria, Apulia South, Sicily.
    \item CLASS 3: Sardinia inland, Sardinia coast, Liguria east, Liguria west.
\end{itemize}
Note that areas in class 2 exactly agree with areas from the South Region. This classification coincides with the separation in classes given by 3-PROP, whereas 2-PROP model grouped together class 1 and class 3. These facts support that our intermediate models have been able to take advantage of  the apparent difference in the structure of the covariance matrices from the South region and the others. When we are looking for a three-class separation, instead of splitting the areas from the Centre-North and Sardinia into these two regions, all Centre-North and Sardinia areas are grouped together, except Umbria, which forms a group alone. Figure \ref{fig:olive3cpc} represents the solution in the principal components of the group Umbria, and we can appreciate the characteristics of this area. The plot corresponding to the second and third variables allows us to see clear differences in some of its principal components. Additionally, we can see that it is also the area with less variability in many directions. In conclusion, a different behavior of the variability in the olive oils from this area seems to be clear. This could be related to the geographical situation of Umbria (the only non-insular and non-coastal area under consideration).

\section{Conclusions and further directions}\label{sec:conclusion}

Cluster analysis of structured data opens up interesting research prospects. This fact is widely known and used in applications where the data themselves share some common structure, and thus clustering techniques are a key tool in functional data analysis. More recently, the underlying structures of the data have increased in complexity, leading, for example, to consider probability distributions as data, and to use innovative metrics, such as earth-mover or Wasserstein distances. This configuration has been used in cluster analysis, for example, in \cite{Barrio}, from a classical  perspective, but also including new perspectives: meta-analysis of procedures, aggregation facilities.... Nevertheless,  to the best of our knowledge, this is the first occasion in which a clustering procedure is used as a selection (of an intermediate model) step in an estimation problem. Our proposal allows improvements in the estimation process and, arguably, often a gain in the interpretability of the estimation thanks to the chosen framework: Classification through the Gaussian Mixture Model. \\

The presented methodology enhances the so-called parsimonious model leading to the inclusion of  intermediate models. They are linked to geometrical considerations on the ellipsoids associated to the covariance matrices of the underlying populations that compose the mixture. These considerations are precisely the essence of the parsimonious model.
The intermediate models arise from clustering covariance matrices, considered as structured data, and using a similarity measure based in the likelihood. The consideration of clustering these objects through other similarities could be appropriate  looking for tools for different goals. In particular, we emphasize on the possibility of clustering based on metrics like the Bures-Wasserstein distance. The role played here by the BIC would have to be tested in the corresponding configurations or, alternatively, replaced by appropriate penalties for choosing between other hierarchical models. Feasibility of the proposal is an essential requirement for a serious essay of a statistical tool. The  algorithms considered in the paper are simple adaptations of Classification Expectation Maximization algorithm, but we think that they could be still improved. We will pursuit on this challenge, looking also for   feasible computations for similarities associated to new pre-established objectives.\\

In summary, through the paper we  have used clustering to explore similarities between groups according to predetermined patterns. In this wider setup, clustering is not a goal in itself, it can be an important tool for specialized analyses.

\subsection*{Supplementary material}

 Github repository containing the 
 R scripts with the algorithms and workflow necessary to reproduce the results of this work. Simulation data of the examples are also included. 
(\url{https://github.com/rvitores/ImprovingModelChoice})

\subsection*{Competing interest}
The authors have no relevant financial or non-financial interests to disclose.

\bibliographystyle{abbrvnat}
\bibliography{bibliography}

\clearpage

\appendix
\label{appendix}

\section{Theoretical results}\label{app:teor}

In this section we are going to further on the comments of Remark \ref{remark}. Given a parsimonious model $\mathscr M$ and fixed values of ${K},G,c_{vol}$ and $c_{sh}$, the problem consists in maximizing the function (\ref{eq:teor}) in $\Theta_{c_{vol},c_{sh}}^{\mathscr M,G}$, the set of parameters $\pmb{\pi},\pmb{\mu},{\pmb{u}} ,\pmb{V},\pmb{C}$ associated with the  clustering model $G$-$\mathscr M$ 
 verifying the size and shape constraints (\ref{eq:shape}). Using the same notation as in \cite{detShape}, denote
\begin{align*}
    \Theta_{c_{vol},c_{sh}} = \left\lbrace (\pmb{\pi},\pmb{\mu},\pmb{\Sigma}) \in [0,1]^{{K}}\times \mathbb R^{d{K}}\times (\mathbb S_{>0}^d)^{{K}} \right. \nonumber \\ 
    \left. \text{verifying the constraints} \  (\ref{eq:shape}) \right \rbrace
\end{align*} 
where $\mathbb S_{>0}^d$ is the set of positive definite symmetric real matrices. If we define the map
\begin{equation*} \begin{array}{ccc}
     \Theta_{c_{vol},c_{sh}}^{\mathscr M,G} & \overset{T}{\longrightarrow} &[0,1]^{{K}}\times \mathbb R^{d{K}}\times (\mathbb S_{>0}^d)^{{K}} \\ 
    \Bigl(\pmb{\pi},\pmb{\mu},{\pmb{u}} ,\pmb{V},\pmb{C}\Bigr)& \longmapsto& \Bigl(\pmb{\pi},\pmb{\mu},\pmb{\Sigma}\bigl({\pmb{u}} ,\pmb{V},\pmb{C}\bigr) \Bigr)
\end{array}  \ ,\end{equation*}
where $\pmb{\Sigma}\bigl({\pmb{u}} ,\pmb{V},\pmb{C}\bigr)$ is the collection of ${{K}}$ covariance matrices created from the parameters ${\pmb{u}} ,\pmb{V},\pmb{C}$,  it is obvious that  $T( \Theta_{c_{vol},c_{sh}}^{\mathscr M,G} )\subset \Theta_{c_{vol},c_{sh}}$.  This and Lemma 1 in \cite{detShape} allow us to replicate the proofs of Proposition 1 and Proposition 2 in \cite{avoiding} to prove the following theorems on  the existence and consistence of the solutions. 

\begin{theorem}\label{theorem1}
    If $\mathbb P$ is a probability that is not concentrated on $K$ points, and $\operatorname{E}_{\mathbb P}\vert\vert\cdot \vert\vert^2<\infty $, the maximum of (\ref{eq:teor}) is achieved at some  $(\pmb{\hat\pi},\pmb{\hat\mu},\pmb{{\hat u}} ,\pmb{\hat V},\pmb{\hat C} )\in \Theta_{c_{vol},c_{sh}}^{\mathscr M,G}$.
\end{theorem}

Given $\lbrace y_{{i}} \rbrace_{{i}=1}^\infty$ independent observations of the distribution ${\mathbb P}$, for each $N$ we can define the empirical distribution ${\mathbb P}_N=(1 / N)\sum_{{i}=1}^N \delta_{\left\{y_{{i}}\right\}}$. The sample problem of maximizing (\ref{eq:teor}) under the constraint (\ref{eq:shape}) coincides with the distributional problem presented here, when we take the probability ${\mathbb P}_N$. Therefore, Theorem \ref{theorem1} also guarantees the existence of the solution of the empirical problem { corresponding to large enough samples drawn from an absolutely continuous distribution.}\\

We use the notation $\theta_0$ for any constrained maximizer of
the theoretical problem for the underlying distribution $\mathbb P$, and
let
\begin{equation*}\theta_n= \bigl(\pmb{\pi^n},\pmb{\mu^n},\pmb{{u}^n} ,\pmb{V^n},\pmb{C^n}\bigr) \end{equation*}
be a sequence of empirical solutions for the sequence of empirical
sample distributions $\lbrace \mathbb P_N \rbrace_{N=1}^\infty$. The following result
states consistency under similar assumptions as in Theorem
\ref{theorem1} if the maximizer of the theoretical problem is assumed to
be unique.
\begin{theorem}
    Let us assume that $\mathbb P$ is not concentrated on $K$
points, $\operatorname{E}_{\mathbb P}\vert\vert\cdot \vert\vert ^2<\infty $ and that $\theta_0 \in  \Theta_{c_{vol},c_{sh}}^{\mathscr M,G} $ is the unique
constrained maximizer of (\ref{eq:teor}) for $ \mathbb P$. If $\lbrace \theta_n \rbrace_{n=1}^\infty$ is a sequence of empirical maximizers of (\ref{eq:teor}) with $\theta_n \in  \Theta_{c_{vol},c_{sh}}^{\mathscr M,G}$, then 
$\theta_n \underset{ }{\longrightarrow} \theta_0 $ almost surely.
\end{theorem}

\renewcommand{\thefigure}{B\arabic{figure}}
\setcounter{figure}{0}

\renewcommand{\thetable}{B\arabic{table}}
\setcounter{table}{0}

\section{Additional simulations} \label{app:additionalSim}

At the suggestion of a reviewer, we present two additional simulation examples that reinforce the ideas presented in Section \ref{sec:cluster}. For the sake of brevity, we only give the results  for the more involved clustering problem. We point out two basic ideas. Since we have introduced a broader family of models, model selection will be more challenging than within the fourteen parsimonious models. This is clearly seen in the former example, but with a sufficiently large sample size, BIC is still able to select the true model. In the latter example, we emphasize that our extension of the parsimonious model is not redundant.\\

First, we repeat the two-dimensional simulation experiment described in Section \ref{sec:cluster},  but assuming the VVE model:
\begin{align*}
\Sigma_k = \gamma_k \beta \Lambda_k \beta^T \ , \quad \quad k=1,\ldots,6.
\end{align*}

\begin{figure*}[ht] 
\centering
   \includegraphics[scale=0.6]{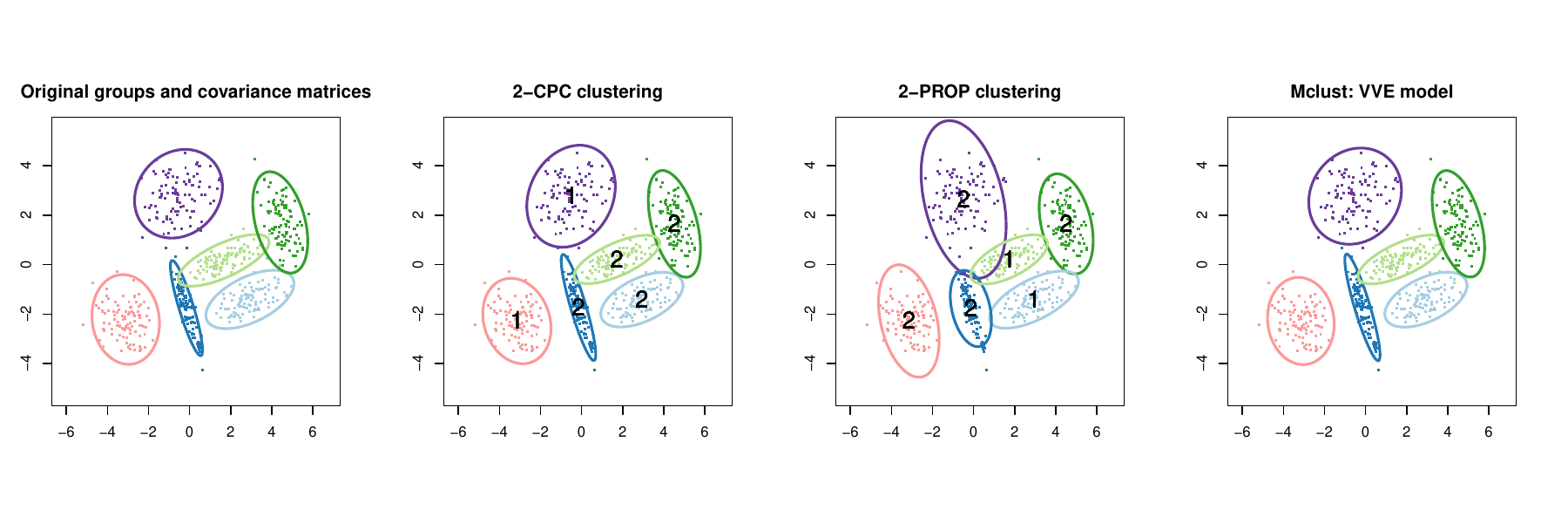}
    \caption{From left to right: 1. Theoretical Gaussian distributions and observations simulated from each distribution in the VVE example. 2. Solution estimated by clustering through 2-CPC model. 3. Solution estimated by clustering through 2-PROP model. 4. VVE clustering solution estimated with $mclust$.}
   \label{fig:A1=A2}
 \end{figure*}

This example allows us to deal with two different situations. The true underlying model verifies the VVE (1-CPC) model, so it also verifies the 2-CPC model, but it does not verify the 2-PROP model. For a sample with $n=50$ observations from each group, we compute the VVE, 2-CPC and 2-PROP solutions for clustering.  Results are shown in Figure \ref{fig:A1=A2}, where we can appreciate that both VVE and 2-CPC models fit the data perfectly, while the constraint of 2-PROP does not allow a good fitting of the data. This is also reflected in Table \ref{table:a1=a2clus}, where the BIC values are computed. The best model in terms of BIC is VVE, but 2-CPC is also competitive. 2-PROP gives much worse BIC values.

 \begin{table}[ht]
\caption{Clustering results in the  VVE example for VVE, 2-CPC and 2-PROP models.}
\medskip
    \centering
    \begin{tabular}{c|ccc}
    \hline
       model & loglik & df &  BIC   \\
      \hline
      \textit{mclust}: VVE & -2116.478 & 30 & \textbf{-4424.864}   \\
      2-CPC &  -2115.219 &  31 &  -4428.742 \\
      2-PROP & -2212.407 &  27 &  -4597.531 \\
          \hline   
    \end{tabular}
    \label{table:a1=a2clus}
\end{table}

Finally, as we did in Table \ref{proportions}, simulations have been repeated 1000 times, for different sample sizes $n$. In each simulation, we are comparing the BIC value obtained for 2-CPC and 2-PROP with the BIC value obtained for the true underlying model VVE.  Results are shown in Table \ref{table:propClust}. 

\begin{table}[ht]
\caption{Proportions of times in which clustering 2-CPC or 2-PROP model improves the model VVE in terms of BIC, for each sample size $n$. }
\medskip
    \centering
    \begin{tabular}{cccc}
        \hline 
        \hspace{0.1cm}Example \hspace{0.1cm}& \hspace{0.1cm}n=50\hspace{0.1cm} & \hspace{0.1cm}n=100\hspace{0.1cm}&\hspace{0.1cm} n=500\hspace{0.1cm} \\
        \hline 
         2-CPC &  0.208  & 0.141 & 0.031 \\
         2-PROP& 0.001 & 0 & 0 \\
         \hline
    \end{tabular}
    \label{table:propClust}
\end{table}
The results are consistent with the ideas set out above. Since 2-PROP model is not verified, the clustering models fitted with this constraint give lower BIC value than VVE. 2-CPC model is verified, it is more flexible than VVE, and the difference in the number of parameters is only one. Thus, this is a rather complicated setting for model selection. Even in this case, if the sample size $n$ is large enough, BIC is able to select the true model in almost all cases.  \\

The second example is similar to the 2-CPC example in \ref{sec:cluster}, but now in dimension $d=10$. We consider $K=6$ distributions, with $G=2$ classes given by
\begin{align*}
       \Sigma_{{k}} &= \gamma_{{k}}\beta_1\Lambda_{{k}} \beta_1^T, \quad {k}=1,2,3, \\
       \Sigma_{{k}} &= \gamma_{{k}}\beta_2\Lambda_{{k}} \beta_2^T ,\quad {k}=4,5,6 \ . 
\end{align*}

Parameters were created so that we get a favorable but not trivial situation for applying clustering algorithms. Figure \ref{fig:simulated} shows a sample created with $n=100$ observations from each group. For this sample, we fit the clustering model 2-CPC, and we compare it with the best model estimated by $mclust$. The results of this simulation are given in Table \ref{tab:simDim10}. 

\begin{table}[ht]
\caption{Clustering results in the 10-dimensional example for the best \textit{mclust} model and 2-CPC.}
\medskip
    \centering
    \begin{tabular}{c|ccc}
    \hline
       model & loglik & df &  BIC   \\
      \hline
      \textit{mclust}: VVE & -7310.521 & 170 & -15708.52  \\
      2-CPC &  -6714.553 &  215 &  \textbf{-14804.45 } \\
          \hline   
    \end{tabular}
    \label{tab:simDim10}
\end{table}

The main advantage of considering our intermediate models against the 14 parsimonious models estimated by $mclust$ in this particular example is that $mclust$ is selecting the model VVE, which it is not exactly verified, because the VVV model involves a substantially larger number of parameters (395 for clustering, 390 for discriminant analysis). This leads to a significant improvement in the BIC value of the 2-CPC model. As a result of this, when we repeated the simulation 1000 times with different sample sizes $n (=50,100,200)$, our model 2-CPC improved in terms of BIC the best model estimated by $mclust$ in 100\% of the simulations, for all the values of $n$ considered.

\section{Algorithms}\label{appendixB}
\subsection{Optimal truncation}

In the algorithms presented, we will repeatedly use the \textit{optimal truncation} algorithm explained in Section 3.1 in \cite{detShape}, which was introduced in \cite{tclust2}. \\

Given $d \geq 0$ and a fixed restriction constant $c \geq 1$, the $m$-truncated value is defined by
\begin{equation*}
d^m= \begin{cases}d & \text { if } d \in[m, c m] \\ m & \text { if } d<m \\ c m & \text { if } d>c m\end{cases} \ .
\end{equation*}

Given $\left\{n_{j}\right\}_{{j}=1}^J \in \mathbb{R}^J_{>0}$ and $\left\{d_{{j} 1}, \ldots, d_{{j} L}\right\}_{{j}=1}^J \in[0, \infty)^{J \times L}$, we define the operator
\begin{equation*}\text {OT}_c\left(\left\{n_{j}\right\}_{{j}=1}^J ;\left\{d_{{j} 1}, \ldots, d_{{j} L}\right\}_{{j}=1}^J\right) \ ,\end{equation*}
which returns $\left\{d_{{j} 1}^*, \ldots, d_{{j} L}^*\right\}_{{j}=1}^J \in[0, \infty)^{J \times L}$ with $d^*_{{j} l}=d_{{j} l}^{m_{opt}}$ for $m_{opt}$ being the optimal threshold value obtained as
\begin{equation*} \label{eq:mopt}
m_{\mathrm{opt}}=\underset{m}{\textnormal{argmin}} \sum_{{j}=1}^J n_{j} \sum_{l=1}^L\left(\log \left(d_{{j} l}^m\right)+\frac{d_{{j} l}}{d_{{j} l}^m}\right) \ .
\end{equation*}

Obtaining that optimal threshold value only requires the maximization of a real-valued function and $m_{opt}$ can be efficiently obtained by performing only $2 \cdot J \cdot L + 1$ evaluations through a procedure which can be fully vectorized  \citep{tclust2}.\\

In the algorithms of the following sections, when working with proportionality models, we will minimize in several situations a function of the type
\begin{align*}\label{eq:PROP}
(\beta,\gamma_1,\ldots,&\gamma_r,\lambda_1, \ldots, \lambda_d) \longmapsto \\
&\sum_{{k}=1}^r\hspace{0.1cm} n_{{k}} \sum_{l=1}^d \Biggl( \log(\gamma_{{k}} \lambda_l)+ \frac{\beta_{l}^T S_{{k}} \beta_l}{\gamma_{{k}} \lambda_{l}}\Biggr) \ ,
\end{align*}
being $\beta$ an orthogonal matrix and $\beta_l,l=1,\ldots,d$ its columns, $\gamma_1,\ldots,\gamma_r$ size parameters verifying the size constraint for $c_{vol}$ and $\lambda_1, \ldots, \lambda_d$ the common shape parameters verifying the shape constraint for $c_{sh}$ and $\prod_{l=1}^d \lambda_l=1$. In this situation, the minimization can be made iteratively, taking into account that:

\begin{itemize}
    \item Fixed the sizes and shapes, the minimization with respect to $\beta$ can be done with the algorithms proposed in \cite{McNicholas1}.
    \item Fixed the orientation and shapes, the optimal unconstrained values of the size are
    \begin{equation*}\gamma_{{k}}^{opt} = \frac{1}{d} \sum_{l=1}^d \frac{\beta_{l}^T S_{{k}} \beta_l}{\lambda_{l}} \quad {k}=1,\ldots,r \ .\end{equation*}
    Therefore, the optimal restricted values for the size are $ \operatorname{OT}_{c_{vol}} \Bigl( \lbrace n_{{k}} \rbrace_{{k}=1}^r ; \lbrace \gamma_{{k}}^{opt} \rbrace_{{k}=1}^r\Bigr)$.
    \item Fixed the orientation and sizes, the optimal unconstrained values of the shapes are:
    \begin{equation*}\lambda_l^{opt} = \frac{1}{N} \sum_{{k}=1}^r n_{{k}} \frac{\beta_{l}^T S_{{k}} \beta_l}{\gamma_{{k}}} \quad l=1,\ldots,d \ .\end{equation*}
    The optimal values verifying the constraint $c_{sh}$ are $ \operatorname{OT}_{c_{sh}} \Bigl( \lbrace 1 \rbrace; \lbrace \lambda_1^{opt},\ldots,\lambda_d^{opt} \rbrace \Bigr)$, and because of the reasoning in Section 3.3 in \cite{detShape}, the optimal values verifying also $\prod_{l=1}^d \lambda_l=1$ are obtained normalizing the result of the optimal truncation operator.  
\end{itemize}

\noindent
When working with CPC models, many times we will come to the conclusion that we have to minimize a slightly different type of function:
\begin{align*}
(\beta,\gamma_1,\ldots&,\gamma_r,\lambda_{1,1}, \ldots, \lambda_{1,d},\ldots, \lambda_{r,d}) \longmapsto \\
&\sum_{{k}=1}^r\hspace{0.1cm} n_{{k}} \sum_{l=1}^d \Biggl( \log(\gamma_{{k}} \lambda_{{k},l})+ \frac{\beta_{l}^T S_{{k}} \beta_l}{\gamma_{{k}} \lambda_{{k},l}}\Biggr) \ .
\end{align*}
In this case, we can repeat analogous comments for the minimization with respect to the sizes and the orientation matrix. For the shape matrices:
\begin{itemize}
     \item Fixed the orientation and sizes, the optimal unconstrained values of the shapes are 
    \begin{equation*}\lambda_{{k},l}^{opt} = \frac{\beta_{l}^T S_{{k}} \beta_l}{\gamma_{{k}}} \quad {k}=1,\ldots,r,\  l=1,\ldots,d \ .\end{equation*}
    For each ${k}=1,\ldots,r$, the optimal values verifying the constraint $c_{sh}$ are the result of the operator $ \operatorname{OT}_{c_{sh}
    } \Bigl( \lbrace 1 \rbrace; \lbrace \lambda_{{k},1}^{opt}\ldots,\lambda_{{k},d}^{opt} \rbrace \Bigr)$, and the optimal values verifying also  $\prod_{l=1}^d \lambda_l=1$ are obtained normalizing the result of that truncation.  
\end{itemize}

\subsection{Classification G-CPC/G-PROP} \label{alg:clas}

In this section we are going to develop the algorithms for the covariance matrices classification models G-CPC and G-PROP minimizing (\ref{eq:CPC1}) and (\ref{eq:PROP1}). Since these algorithms are included in the algorithms for cluster analysis, determinant and shape constraints are also included. When focusing on the original problem of Section \ref{section:clas}, these constraints should be omitted, which can be done taking $c_{vol}=c_{sh}=\infty$. The input of the algorithm is
\begin{align*}
    \textbf{Classification G-CPC/PROP}& \\
    \Bigl(S_1,\ldots,S_{{K}}, n_1,\ldots,n_{{K}},G,c_{sh},&c_{vol} ,nstart_1\Bigr) \ ,
\end{align*}
where $S_1,\ldots,S_{{K}}$ are the sample covariance matrices, $n_1,\ldots,n_{{K}}$ the the sample lengths, $G$ the number of classes, $c_{sh},c_{vol}$ the values of the constants for the determinant and shape constraints and $nstart_1$ the number of random initializations. The parameters of the minimization are ${\pmb{u}}=({u}_1,\ldots,{u}_{{K}})$, $\pmb{\gamma}=(\gamma_1,\ldots,\gamma_{{K}})$ $\pmb{\beta}=(\beta_1,\ldots,\beta_G)$ and $\pmb{\Lambda}=(\Lambda_1,\ldots,\Lambda_s)$, where $s={K}$ in G-CPC and $s=G$ in G-PROP, and they are also the output of the algorithm. A detailed presentation of the algorithm is given as follows:

\begin{enumerate}[left=-0.3cm]
    \item[\textbf{1.}] \textbf{Initialization:} We start taking a random vector of indexes $\pmb{{u}^0} \in \mathscr H$. Then we take:
    \begin{itemize}
        \item[$\pmb{\beta^0:}$] For each $g = 1,\ldots,G$, we take ${k}$ such that ${u}_{{k}}^0=g$, and we define $\beta_g$ as the eigenvectors of $S_{{k}}$.
        \item[$\pmb{\Lambda^0:}$] $\rightarrow$ G-PROP: For each $g=1,\ldots,{K}$, taking the same ${k}$ as before,
        \begin{align*}
        \Lambda_g^0 =& \operatorname{OT}_{c_{sh}}\Bigl( \lbrace 1 \rbrace ;\operatorname{diag}(\beta_{g}^TS_{{k}}\beta_{g})\Bigr), \\ \Lambda_g^0 = & \frac{\Lambda_g^0 }{\operatorname{prod}(\Lambda_g^0 )^{1/d}} \ .
        \end{align*}
        \item[] $\rightarrow$ G-CPC: For each ${k}=1,\ldots,{K}$,
        \begin{align*}  \Lambda_{{k}}^0 =& \operatorname{OT}_{c_{sh}}\Bigl( \lbrace 1 \rbrace ;\operatorname{diag}\bigl( \beta_{{u}_{{k}}^0}^TS_{{k}}\beta_{{u}_{{k}}^0}\bigr)\Bigr),  \\
        \Lambda_{{k}}^0 =& \frac{\Lambda_{{k}}^0 }{\operatorname{prod}(\Lambda_{{k}}^0 )^{1/d}} \ .
        \end{align*}
        \item[$\pmb{\gamma^0:}$] For each ${k}=1,\ldots,{K}$,
    \begin{align*}
    &\rightarrow\textnormal{G-PROP:} \ &\gamma_{{k}}^0 = \frac{1}{d}\operatorname{tr}\Bigl((\Lambda_{{u}_{{k}}^0}^0)^{-1}\beta_{{{u}_{{k}}^0}}^TS_{{k}}\beta_{{{u}_{{k}}^0}}\Bigr) \\
    &\rightarrow\textnormal{G-CPC:}\  &\gamma_{{k}}^0 = \frac{1}{d}\operatorname{tr}\Bigl((\Lambda_{{k}}^0)^{-1}\beta_{{{u}_{{k}}^0}}^TS_{{k}}\beta_{{{u}_{{k}}^0}}\Bigr) \ .
    \end{align*}
    Constrained values: 
    \begin{equation*} (\gamma_1^0,\ldots,\gamma_d^0)=\operatorname{OT}_{c_{vol}} \Bigl( \lbrace n_{{k}} \rbrace_{{k}=1}^{{K}} ; \lbrace \gamma_{{k}}^{0} \rbrace_{{k}=1}^{{K}}\Bigr) \ .\end{equation*}
    \end{itemize}

\item[\textbf{2.}] \textbf{Iterations:} The following steps are repeated until convergence:

\begin{itemize}[left=-0.1cm]
    \item \textbf{\pmb{{u}-V step:}} Based on the current parameters $\pmb{{u}^m},\pmb{\gamma^m},\pmb{\beta^m},\pmb{\Lambda^m}$, we are going to optimize with respect to $\pmb{{u}}$ and the variable parameters of each parsimonious model. The variable parameters will be also optimized in the following step, thus its value will not be updated here. Size parameters $\pmb{\gamma}$ don't affect the selection of the best ${\pmb u}$, thus it is enough to find for each ${k}=1,\ldots,{K}$ the value of  ${u}_{{k}}$ for which taking the common parameters $C_{{u_k}}$ we obtain a lower value in the minimization with respect to the variable parameters of 
    \begin{equation*}
     R(\beta,\Lambda)=\sum_{l=1}^d \frac{\beta_{l}^T S_{{k}} \beta_{l}}{\lambda_{l}} \ .   
    \end{equation*}
    
   $\rightarrow$ G-PROP: The parameters $\pmb \Lambda, \pmb \beta$ are common, we are only minimizing with respect to ${\pmb u}$. For each ${k} = 1,\ldots,{K}$,
        \begin{align*}
           {u}_{{k}}^{m+1} = \underset{g\in \lbrace 1,\ldots,G\rbrace}{\operatorname{{argmin}}}\quad  R(\beta_g^m,\Lambda_g^m) \ .
        \end{align*}
        
        $\rightarrow$ G-CPC: The parameters $\pmb \beta$ are common. For each ${k}=1,\ldots,{K}$,
        \begin{align*} 
        \Tilde\Lambda_{{k},g}=& \operatorname{OT}_{c_{sh}}\Bigl( \lbrace 1 \rbrace ;\operatorname{diag}\bigl( (\beta_{g}^m)^T S_{{k}}\beta_{g}^m\bigr)\Bigr) \ ,  \\
        \Tilde\Lambda_{{k},g}  =& \frac{\Tilde\Lambda_{{k},g} }{\operatorname{prod}(\Tilde\Lambda_{{k},g}  )^{1/d}} \ , \\ {u}_{{k}}^{m+1} =& \underset{g\in \lbrace 1,\ldots,G\rbrace}{\operatorname{{argmin}}}\quad  R(\beta_g^m,\Tilde \Lambda_{{k},g}) \ .\end{align*}

    \item \textbf{\pmb{V-C step:}} Based on the current parameters $\pmb{{u}^{m+1}},\pmb{\gamma^m},\pmb{\beta^m},\pmb{\Lambda^m}$, we are going to optimize with respect to $\pmb{\gamma},\pmb{\beta},\pmb{\Lambda}$. This optimization requires iterations. Setting $s=0$, and considering the initial solutions
    \begin{equation*} \pmb{\bar \gamma^0 }= \pmb{\gamma^{m}} \quad\quad  \pmb{\bar \beta^0 }= \pmb{\beta^{m}} \quad \quad \pmb{\bar \Lambda^0 }= \pmb{\Lambda^{m}} \ , \end{equation*}
   the following steps are repeated until convergence:

    \begin{itemize}
        \item[$\pmb{s:}$] $s= s + 1$
        \item[$\pmb{\bar\gamma^{s}:}$] Update the size parameters.  For each ${k}=1,\ldots,{K}$,\\

        $\rightarrow$ G-PROP: 
    \begin{align*}
        \bar \gamma_{{k}}^{s} =\frac{1}{d}\operatorname{tr}\Bigl((\bar \Lambda_{{u}_{{k}}^{m+1}}^{s-1})^{-1}(\bar \beta_{{u}_{{k}}^{m+1}}^{s-1})^TS_{{k}}\bar\beta_{{u}_{{k}}^{m+1}}^{s-1}\Bigr) \ .
    \end{align*}
    $\rightarrow$ G-CPC:
    \begin{align*}
        \bar \gamma_{{k}}^{s} = \frac{1}{d}\operatorname{tr}\Bigl((\bar \Lambda_{{k}}^{s-1})^{-1}(\bar\beta_{{u}_{{k}}^{m+1}}^{s-1})^T S_{{k}}\bar\beta_{{u}_{{k}}^{m+1}}^{s-1}\Bigr) \ . 
    \end{align*}  
    
    Then we apply the size constraint:
    \begin{equation*} \pmb{\bar\gamma^{s}}=\operatorname{OT}_{c_{vol}} \Bigl( \lbrace n_{{k}} \rbrace_{{k}=1}^r ; \lbrace \bar\gamma_{{k}}^s \rbrace_{{k}=1}^{{K}}\Bigr) \ .\end{equation*}

    \item[$\pmb{\bar\Lambda^{s}:}$] Update the shape  parameters.\\
    
     $\rightarrow$ G-PROP: For each $g=1,\ldots,G$,
         \begin{align*}
         {\bar\Lambda_g^{s}} &=\operatorname{OT}_{c_{sh}}\Biggl( \lbrace 1 \rbrace ; \\
         &\operatorname{diag}\Biggl(\frac{1}{N} \sum_{{k}:{u}_{{k}}^{m+1}=g} n_{{k}} \frac{(\bar \beta_{g}^{s-1})^T S_{{k}} \bar \beta_g^{s-1}}{\bar\gamma_{{k}}^s}\Biggr) \Biggr), \\
           {\bar\Lambda_g^{s}}&= \frac{{\bar\Lambda_g^{s}}}{\operatorname{prod}({\bar\Lambda_g^{s}})^{1/d}} \ .
           \end{align*}
           
        $\rightarrow$ G-CPC: For each ${k}=1,\ldots,{K}$,
  \begin{align*}
    {\bar\Lambda_{{k}}^{s}} =& \operatorname{OT}_{c_{sh}}\left( \lbrace 1 \rbrace ; \operatorname{diag}\Biggl(\frac{(\bar \beta_{{u_k}^{m+1}}^{s-1})^T S_{{k}} \bar \beta_{{u_k}^{m+1}}^{s-1}}{\bar\gamma_{{k}}^s} \Biggr) \right) , \\
   {\bar\Lambda_{{k}}^{s}}=& \frac{{\bar\Lambda_{{k}}^{s}}}{\operatorname{det}({\bar\Lambda_{{k}}^{s}})^{1/d}} .
  \end{align*}

    \item[$\pmb{\bar\beta^{s}:}$] Update the rotation  parameters. For each $g=1,\ldots,G$, the algorithms in \cite{McNicholas1} allow us to find, for each $g=1,\ldots,G$, a rotation matrix $\bar\beta_g^s$ minimizing: 
        \begin{align*}
        &\rightarrow\textnormal{G-PROP:} \ \beta \mapsto \sum_{{k}: {u}_{{k}}^{m+1}=g}n_{{k}} \hspace{0.1cm}\sum_{l=1}^d \frac{\beta_{l}^T S_{{k}} \beta_l}{\bar\gamma_{{k}}^s \bar\lambda_{g,l}^s} \ .\\
         &\rightarrow \textnormal{G-CPC:} \quad \beta \mapsto \sum_{{k}: {u}_{{k}}^{m+1}=g} n_{{k}} \hspace{0.1cm}\sum_{l=1}^d \frac{\beta_{l}^T S_{{k}} \beta_l}{\bar\gamma_{{k}}^s \bar\lambda_{{k},l}^s} \ .    
        \end{align*} 

    \end{itemize}
Once the iterations have finished, we update the parameters
\begin{equation*} \pmb{\gamma^{m+1} }= \pmb{\bar \gamma^{s}} \quad\quad  \pmb{ \beta^{m+1} }= \pmb{\bar\beta^{s}} \quad \quad \pmb{ \Lambda^{m+1} }= \pmb{\bar \Lambda^{s}} \ .\end{equation*}

\end{itemize}
 
\item[\textbf{3.}] \textbf{Evaluate the target function}: Steps 1 and 2 are repeated $nstart_1$ times. At each step, we evaluate the target function (\ref{eq:CPC1}) or (\ref{eq:PROP1}), and we keep the parameters estimated in the iteration with the best value of the target function.
\end{enumerate}

\subsection{Clustering G-CPC/G-PROP} \label{alg:cluster}

In this section we are going to give a detailed explanation of the algorithms for model-based clustering G-CPC and G-PROP presented in Section \ref{sec:cluster} for minimizing (\ref{eq:verG}). The algorithms for fitting the corresponding discriminant analysis models can be easily deduced from these. The input of the clustering algorithm is:
\begin{align*}
    \textbf{clustering G-CPC/PROP} &\\
    \Bigl(X,G,{K},c_{sh},c_{vol},&nstart_1,nstart_2\Bigr) \ ,
\end{align*}
where $X$ is the matrix with $N$ observations of $d$ variables, $G$ is the number of classes, ${{K}}$ is the number of clusters, $c_{sh},c_{vol}$ are the values for the determinant and shape constraints, $nstart_1$ is the number of random initializations in the classification algorithm, and $nstart_2$ is the number of random initialization in the clustering algorithm. The parameters of the minimization are $\pmb{\pi}=(\pi_1,\ldots,\pi_{{K}})$, $\pmb{\mu}=(\mu_1,\ldots,\mu_{{K}})$, ${\pmb{u}}=({u}_1,\ldots,{u}_{{K}})$, $\pmb{\gamma}=(\gamma_1,\ldots,\gamma_{{K}})$ $\pmb{\beta}=(\beta_1,\ldots,\beta_G)$ and $\pmb{\Lambda}=(\Lambda_1,\ldots,\Lambda_s)$, where $s={K}$ in G-CPC and $s=G$ in G-PROP. A detailed presentation of the algorithm is given as follows:

\begin{enumerate}[left=-0.3cm]
 \item[\textbf{1.}] \textbf{Initialization:} We start taking a random vector of indexes $\pmb{{u}^0}\in \mathscr H$. Then we take:
    \begin{itemize}
        \item[$\pmb{\pi^0:}$] Equal weights: $\pi_{{k}}^0=\frac{1}{{K}} \quad {k}=1,\ldots, {{K}}\ .$
        \item[$\pmb{\mu^0:}$] Denote by $(\bar \mu_1,\ldots, \bar \mu_{{K}})$ the solution obtained by the $R$ function $tclust$ \citep{tclust,tclust3} for a random sample of length $N/2$ of $X$, number of groups ${{K}}$, eigenvalue constraint given by $c=\min\lbrace c_{sh},c_{vol}\rbrace$ and a suitable number of starts. We are considering as initial solution a random perturbation of the values obtained. If $S=\operatorname{cov}(X)$, we are considering
        \begin{equation*}\mu_{{k}}^0 = \bar \mu_{{k}} + \frac{1}{10} N(0,S)\quad \quad {k}=1,\ldots, {{K}}\ .\end{equation*} 
        \item[$\pmb{\beta^0:}$] $\beta_g^0 = I_d, \quad g = 1,\ldots,G\ .$
        \item[$\pmb{\Lambda^0:}$] $\Lambda_{{k}}^0 = (1,\ldots,1),\quad$ ${k}=1,\ldots,G$ in G-PROP, and ${k}=1,\ldots,{K}$ in G-CPC.
        \item[$\pmb{\gamma^0:}$] $\gamma_{{k}}^0=1, \quad {k}=1,\ldots,{K}\ .$ \\
        
        (This simple initial solution verifies determinant and shape constraint independently of the $c_{sh}$ and $c_{vol}$ values)
\end{itemize}
\item[\textbf{2.}] \textbf{Iterations:} The E and M steps are repeated until convergence:

\begin{itemize}[left=-0.1cm]
    \item \textbf{E step}: Given the current values of the parameters $\pmb{\pi^{m}},\pmb{\mu^{m}},\pmb{{u}^{m}}$, $\pmb{V^{m}},\pmb{C^{m}}$, we compute the posterior probabilities
    \begin{equation*} 
        z_{{i},{{k}}} = \frac{\pi_{{k}}\thinspace \phi\Bigl(y_{{i}}\Big \vert\mu_{{k}}^m,\Sigma_{{k}}^m\Bigr)}{\sum_{l=1}^{{K}} \pi_l\thinspace \phi\Bigl(y_{{i}}\vert\mu_l^m,\Sigma_{{l}}^m\Bigr)}
    \end{equation*}
    for ${k}=1,\ldots, {{K}}\quad {{i}}=1,\ldots,N$, where the matrix $\Sigma_{{k}}^m$ is defined by 
    \begin{align*}
        &\rightarrow \textnormal{G-PROP:} \ \Sigma_{{k}}^m=\gamma_{{k}}^m \beta_{{u}_{{k}}^m} \Lambda_{{u}_{{k}}^m}^m\beta_{{u}_{{k}}^m}^T . \\
        &\rightarrow \textnormal{G-CPC:}\quad  \Sigma_{{k}}^m=\gamma_{{k}}^m \beta_{{u}_{{k}}^m} \Lambda_{{k}}^m\beta_{{u}_{{k}}^m}^T .
     \end{align*}
    \item \textbf{M step}: In this step, we have to maximize the complete log-likelihood (\ref{eq:verCompleta}) given the expected values $\lbrace z_{{i},{{k}}} \rbrace_{{i},{{k}}}$. 

    \begin{itemize}
        \item[$\pmb{n:}$] $n_{{k}}= \sum_{{{i}}=1}^N z_{{i},{{k}}} \quad {k}=1,\ldots, {{K}}\ .$
        \item[$\pmb{\pi^{m+1}:}$] $\pi_{{k}}^{m+1}=\frac{n_{{k}}}{N} \quad {k}=1,\ldots, {{K}}\ .$
        \item[$\pmb{\mu^{m+1}:}$] $\mu_{{k}}^{m+1}=\frac{\sum_{{{i}}=1}^N z_{{i},{{k}}}y_{{i}}}{n_{{k}}} \quad {k}=1,\ldots, {{K}}\ .$
         \item[$\pmb{S:}$] For ${k}=1,\ldots,{K}:$
         \begin{equation*}
           S_{{k}}=\frac{1}{n_{{k}}}\sum_{{{i}}=1}^N z_{{i},{{k}}} (y_{{i}}-\mu_{{k}}^{m+1})(y_{{i}}-\mu_{{k}}^{m+1})^T \ .  
         \end{equation*}
        \item[$\pmb{\textnormal{Class.}:}$] We solve the covariance matrix classification problem for the computed values:
    \end{itemize}
\end{itemize}
\begin{align*}
&(\pmb{{u}^{m+1}},\pmb{\gamma^{m+1}},\pmb{\Lambda^{m+1}},\pmb{\beta^{m+1}})= \\
    &\textbf{Classification G-CPC/PROP} \\
& \quad \Bigl(S_1,\ldots,S_{{K}},n_1,\ldots,n_{{K}},G,c_{sh},c_{vol},nstart_1 \Bigr) \ .
\end{align*}

\item[\textbf{3.}] \textbf{Evaluate the target function}: Steps 1 and 2 are repeated $nstart_2$ times. At the end of each different initialization, we evaluate the target function (\ref{eq:verG}), and we keep the parameters estimated in the iteration with the best value of the target function.

\end{enumerate}

\renewcommand{\thefigure}{D\arabic{figure}}
\setcounter{figure}{0}
\section{Supplementary figures}\label{app:fig}

This appendix includes illustrative graphs for the best models, in terms of BIC, fitted in the last three real data examples presented in the paper ``Improving Model Choice in Classification: An Approach Based on Clustering of Covariance Matrices'':
\begin{itemize}
    \item Crabs data set: 2-PROP discriminant analysis model.
    \item Cancer data set: 3-CPC clustering.
    \item Olive oil data set: 3-CPC discriminant analysis.
\end{itemize}

We also include a graph of the simulation data used in the second example of Appendix B. Each figure is accompanied by the corresponding explanatory caption. For a better understanding of the models, graphics should be enlarged, in order to distinguish the different colors and shapes. 
The ellipses in the figures show the contours of the fitted multivariate component densities. While the contours of proportional matrices in a lower dimensional subspace are still proportional, it is important to notice that the analogous property is not true for Common Principal Components, unless the subspace is generated by a subset of the Common Principal Components. For this reason, ellipses representing the contours of the densities verifying the Common Principal Component hypotheses might not preserve common directions. \\

\begin{landscape}
\thispagestyle{empty}
    \begin{figure}[ht] 
        \begin{center}
             \textbf{\Large CRABS DATA SET: Discriminant Analysis 2-PROP model}\par\bigskip
        \end{center}
        
        \hspace{-1cm}     
       \includegraphics[scale=0.31,angle=0]{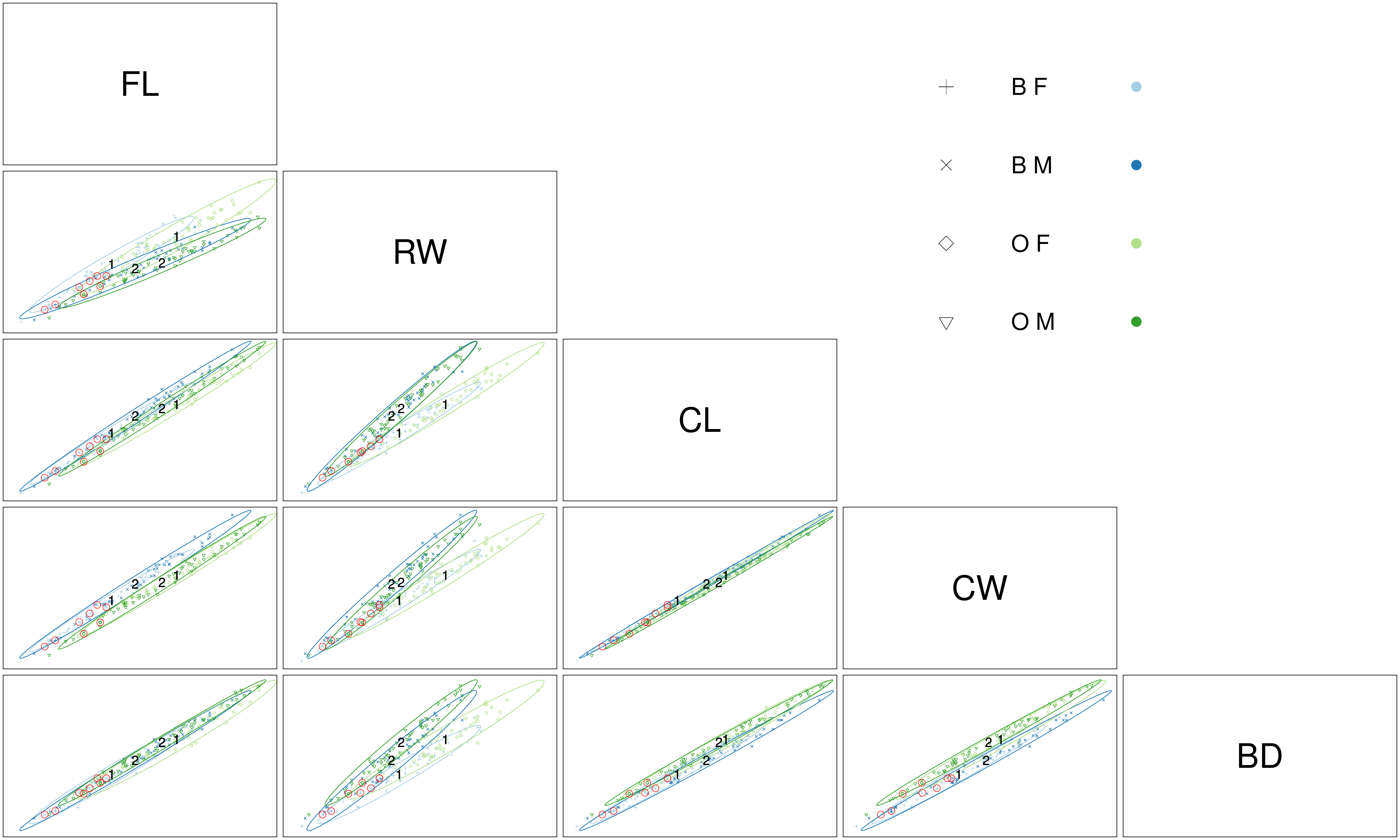}
       \caption{Discriminant analysis obtained from 2-PROP model in the Crabs data set. Color represents the classification created by the model, and the associated ellipses are the contours of the estimated mixture densities, grouped into the classes given by indexes in black. Point shapes represent the original groups. Misclassified observations are marked with red circles.}
       \label{fig:crabs}
     \end{figure}

\clearpage 
\thispagestyle{empty}
 \begin{figure}[ht] 
    \begin{center}
             \textbf{\Large CANCER DATA SET: Clustering 3-CPC model}\par\bigskip
        \end{center}
            \hspace{-1cm}
       \includegraphics[scale=0.21,angle=0]{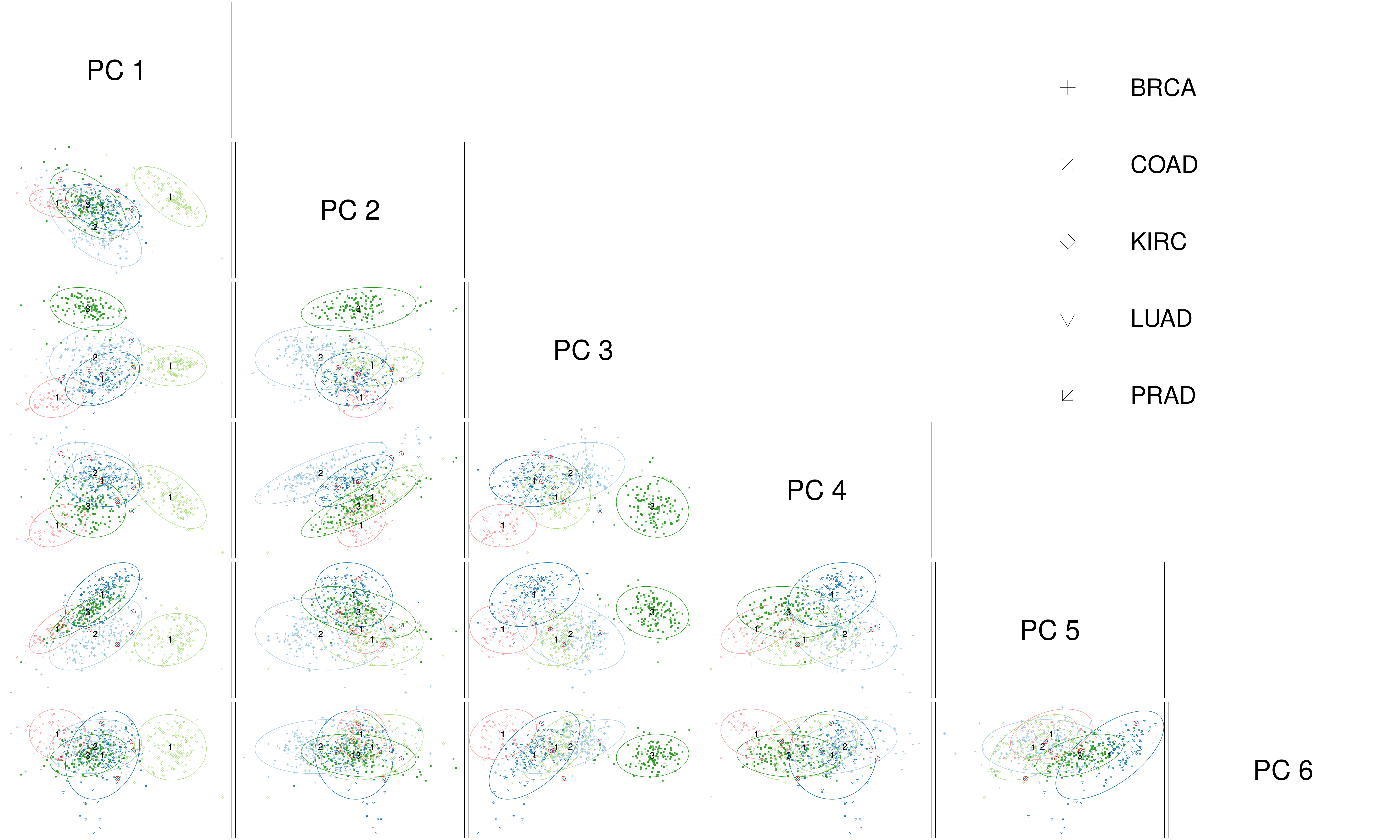}
       \caption{Clustering obtained from 3-CPC model in Gene expression cancer data set. Color represents the clusters created by the model, and the associated ellipses are the contours of the estimated mixture densities, grouped into the classes given by indexes in black. Point shapes represent the original groups. Observations lying on different clusters from the originals are marked with red circles.}
       \label{fig:cancer3cpc}
\end{figure}

\vspace{3cm}
\clearpage 
\thispagestyle{empty}
 \begin{figure}[ht] 
        \begin{center}
             \textbf{\Large OLIVE OIL DATA SET: Discriminant Analysis 3-CPC model}\par\bigskip
        \end{center}
            \hspace{-1cm}
       \includegraphics[scale=0.21,angle=0]{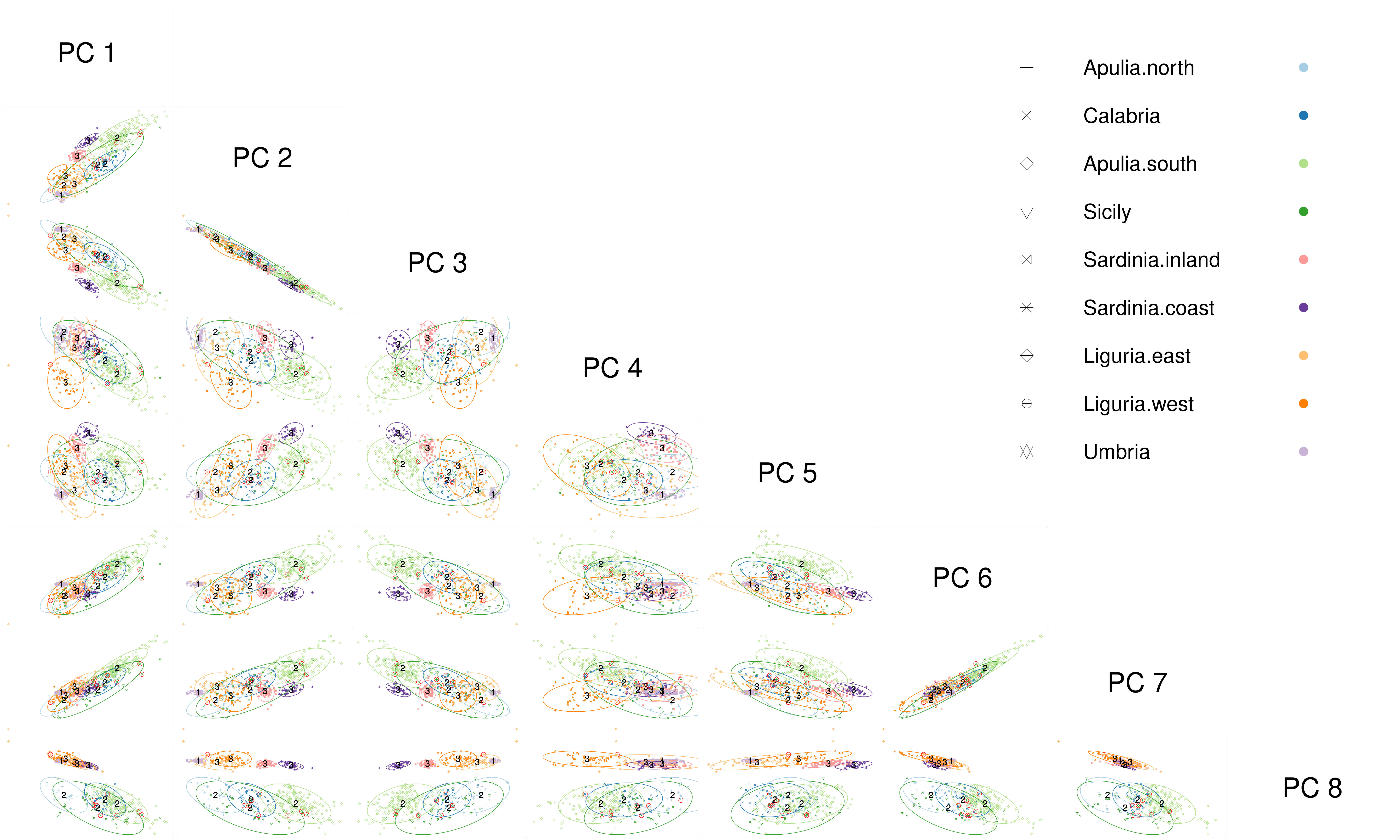}
       \caption{Discriminant analysis obtained from 3-CPC model in the Italian Olive Oil data set.  Color represents the classification created by the model,  and the associated ellipses are the contours of the estimated mixture densities, grouped into the classes given by indexes in black. Point shapes represent the original groups. Misclassified observations are marked with red circles.}
       \label{fig:olive3cpc}
\end{figure}

\vspace{3cm}
\clearpage 
\thispagestyle{empty}
 \begin{figure}[ht] 
        \begin{center}
             \textbf{\Large 2-CPC SIMULATED DATA: Original groups}\par\bigskip
        \end{center}
            \hspace{-0.1cm}
       \includegraphics[scale=0.46,angle=0]{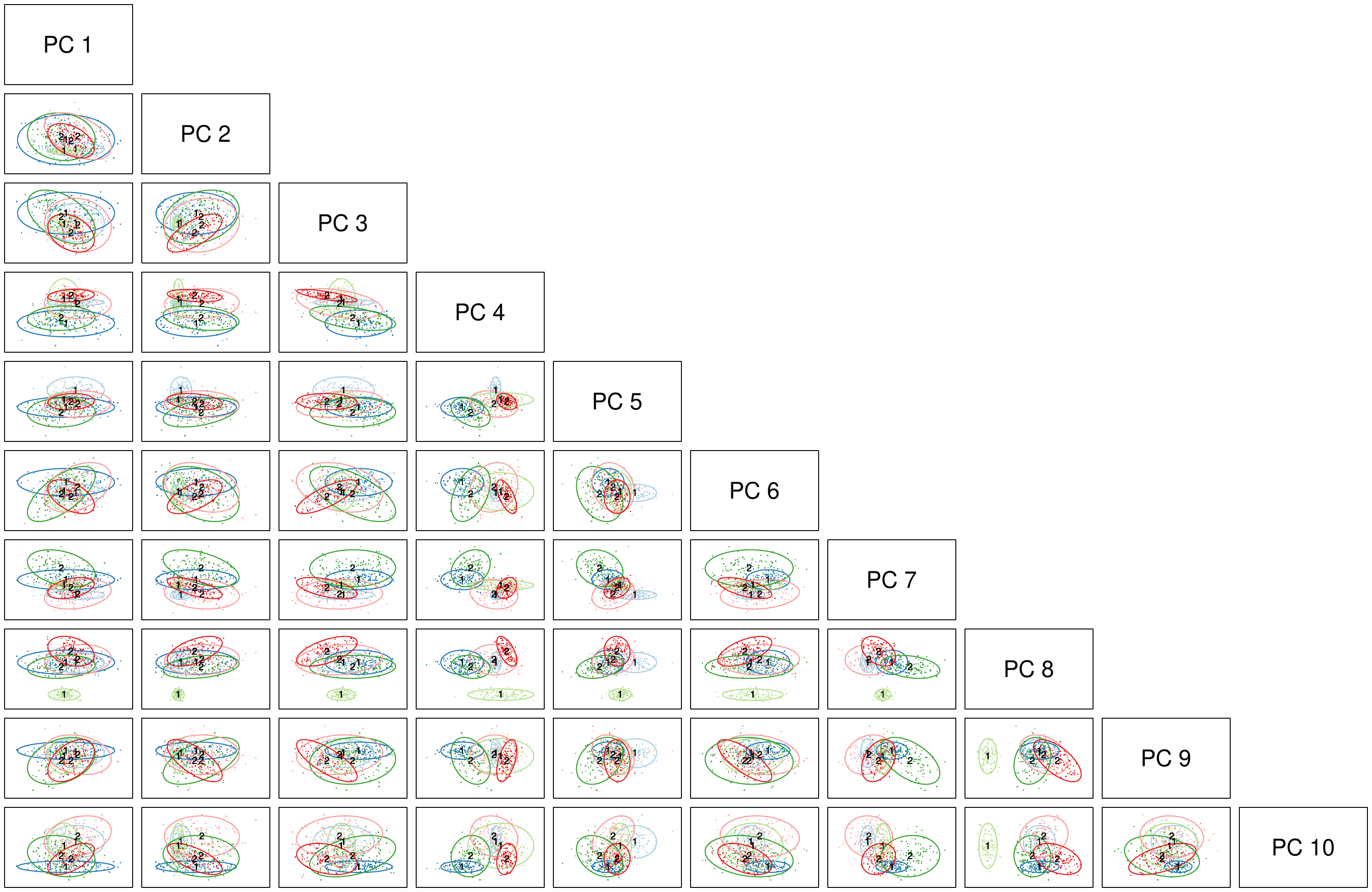}
       \caption{Data simulated from the 2-CPC model in the ten-dimensional example presented in Appendix B, represented in the common principal components of the first class. Color represents the original groups, and the associated ellipses are the contours of the true mixture densities, grouped into the classes given by indexes in black. }
       \label{fig:simulated}
\end{figure}

\end{landscape}

\end{document}